\RequirePackage{fix-cm} 
\documentclass[a4paper, twoside, reqno, dvips, 12pt]{amsart}
\usepackage{fixltx2e}   

\usepackage{etex}

\usepackage[latin1]{inputenc}
\usepackage[T1]{fontenc}

\usepackage{esint}
\usepackage{dsfont}
\usepackage{xspace}
\usepackage{amsbsy}
\usepackage{amsgen}
\usepackage{amsthm}
\usepackage{amssymb}
\usepackage{amsmath}
\usepackage{prodint}
\usepackage{wasysym}
\usepackage{upgreek}
\usepackage{amsfonts}
\usepackage{stmaryrd}
\usepackage{mathtools}

\usepackage[mathcal, mathscr]{euscript}

\usepackage{mathrsfs}
\DeclareMathAlphabet{\mathscrbf}{OMS}{mdugm}{b}{n}

\usepackage{a4wide}

\usepackage[dvipsnames, table]{xcolor}
\definecolor{bckg}{RGB}{20.8, 20.8, 20.8}
\definecolor{oneblue}{rgb}{0.0, 0.0, 0.85}
\definecolor{Lightblue}{RGB}{214, 214, 214}
\definecolor{bluepigment}{rgb}{0.2, 0.2, 0.6}
\definecolor{charcoal}{rgb}{0.21, 0.27, 0.31}
\definecolor{denimblue}{rgb}{0.08, 0.38, 0.74}
\definecolor{Lightgray}{rgb}{0.89, 0.89, 0.89}
\definecolor{darkgrey}{rgb}{0.273, 0.281, 0.30}
\definecolor{darkelectricblue}{rgb}{0.33, 0.41, 0.47}

\usepackage[sort&compress, comma, square, numbers]{natbib}

\usepackage{psfrag}
\usepackage{graphicx}
\usepackage{subfigure}
\usepackage{morefloats}
\usepackage{indentfirst}

\usepackage{acronym}
\usepackage{microtype}
\usepackage[labelsep=period,%
            labelfont={bf,sf,color=bluepigment},%
            justification=raggedright]{caption}

\usepackage[perpage, symbol]{footmisc}

\usepackage[usenames, dvipsnames, pdf]{pstricks}
\usepackage{epsfig}
\usepackage{pst-grad} 
\usepackage{pst-plot} 

\usepackage[colorlinks,
           urlcolor=oneblue,
           linkcolor=denimblue,
           citecolor=NavyBlue,
           bookmarksopen=false,
           pdfpagemode=UseNone,
           pagebackref]{hyperref}

\usepackage[explicit]{titlesec}

\titleformat{\section}[block]
  {\color{NavyBlue}\Large\sffamily\bfseries}
  {}
  {0.0em}
  {\colorbox{bckg!5}{\strut\parbox{\dimexpr\linewidth-2\fboxsep\relax}{\thesection. #1}}}
  [\vspace*{0.33em}]

\titleformat{name=\section,numberless}[block]
  {\color{NavyBlue}\Large\sffamily\bfseries}
  {}
  {0.0em}
  {\colorbox{bckg!5}{\strut\parbox{\dimexpr\linewidth-2\fboxsep\relax}{#1}}}
  [\vspace*{0.33em}]

\titleformat{\subsection}
  {\color{NavyBlue}\large\sffamily\bfseries}
  {}
  {0.0em}
  {\colorbox{bckg!5}{\parbox{\dimexpr\linewidth-2\fboxsep\relax}{\thesubsection. #1}}}
  [\vspace*{0.33em}]

\titleformat{name=\subsection,numberless}
  {\color{NavyBlue}\Large\sffamily\bfseries}
  {}
  {0em}
  {\colorbox{bckg!5}{\parbox{\dimexpr\linewidth-2\fboxsep\relax}{#1}}}
  [\vspace*{0.33em}]

\titleformat{\subsubsection}
  {\color{bluepigment}\sffamily\normalsize\bfseries}
  {\thesubsubsection}
  {0.5em}
  {#1}
  [\vspace*{0.33em}]

\titleformat{\paragraph}[runin]
  {\color{bluepigment}\sffamily\small\bfseries}
  {}
  {0em}
  {#1}

\titlespacing{\section}{0.0em}{1.5em plus 2pt minus 2pt}%
{1.0em plus 2pt minus 2pt}[0em]
\titlespacing{\subsection}{0.5em}{1.5em plus 2pt minus 2pt}%
{1.0em}[0em]
\titlespacing{\subsubsection}{0.5em}{1.5em plus 2pt minus 2pt}%
{1.0em plus 2pt minus 2pt}[0em]

\usepackage{titletoc}

\contentsmargin{0.5em}
\newlength{\tocsep} 
\titlecontents{section}[\tocsep]
  {\addvspace{10pt}\bfseries\sffamily}
  {\contentslabel[\thecontentslabel]{\tocsep}}
  {}
  {\ \titlerule*[0.75pc]{.}\ \thecontentspage}
  []
\titlecontents{subsection}[\tocsep]
  {\addvspace{8pt}\sffamily}
  {\contentslabel[\thecontentslabel]{\tocsep}}
  {}
  {\ \titlerule*[0.5pc]{.}\ \thecontentspage}
  []
\titlecontents*{subsubsection}[\tocsep]
  {\addvspace{2pt}\footnotesize\sffamily}
  {}
  {}
  {\ \titlerule*[0.35pc]{.}\ \thecontentspage}
  [\\*]

\makeatletter
\def\@setauthors{%
  \begingroup
  \def\thanks{\protect\thanks@warning}%
  \trivlist
  \centering\footnotesize \@topsep30\p@\relax
  \advance\@topsep by -\baselineskip
  \item\relax
  \author@andify\authors
  \def\\{\protect\linebreak}%
  \textsc{\normalsize\textcolor{darkelectricblue}{\authors}}%
  \ifx\@empty\contribs
  \else
    ,\penalty-3 \space \@setcontribs
    \@closetoccontribs
  \fi
  \endtrivlist
  \endgroup
}
\def\@settitle{\begin{center}%
  \baselineskip14\p@\relax
    \bfseries
    \textsc{\Large\textcolor{charcoal}{\@title}}
  \end{center}%
}
\makeatother

\usepackage{enumitem}
\setlist[description]{%
  topsep=30pt,               
  itemsep=5pt,               
  font={\bfseries\sffamily\color{NavyBlue}}, 
}

\usepackage{fancyhdr}
\usepackage{lastpage}

\newcommand*\Title{\textcolor{bluepigment}{Serre-type equations in deep water}}
\newcommand*\Authors{\textcolor{bluepigment}{D.~Dutykh, D.~Clamond \& M.~Chhay}}
\newcommand*{\plogo}{\textcolor{gray}{{\texttt{arXiv.org} / \textsc{hal}}}} 

\numberwithin{equation}{section}

\newtheorem{remark}{Remark}

\renewcommand{\mu}{\upmu}
\renewcommand{\nu}{\upnu}

\newcommand{\M}{\mathds{M}}

\newcommand{\R}{\mathds{R}}

\renewcommand{\phi}{\upphi}
\newcommand{\vt}{\tilde{v}}
\newcommand{\Dd}{\mathds{D}}
\newcommand{\Id}{\mathds{I}}

\newcommand{\Km}{\mathds{K}}
\newcommand{\Ss}{\mathds{S}}
\newcommand{\ud}{\mathrm{d}}
\newcommand{\ue}{\mathrm{e}}
\newcommand{\ui}{\mathrm{i}}
\renewcommand{\beta}{\upbeta}
\renewcommand{\L}{\mathcal{L}}
\renewcommand{\H}{\mathcal{H}}
\renewcommand{\S}{\mathcal{S}}

\newcommand{\eps}{\varepsilon}
\renewcommand{\alpha}{\upalpha}
\renewcommand{\gamma}{\upgamma}
\renewcommand{\kappa}{\upkappa}

\newcommand{\g}{\boldsymbol{g}}
\newcommand{\n}{\boldsymbol{n}}
\newcommand{\x}{\boldsymbol{x}}
\newcommand{\z}{\boldsymbol{z}}
\renewcommand{\k}{\boldsymbol{k}}
\renewcommand{\u}{\boldsymbol{u}}
\renewcommand{\lambda}{\uplambda}
\newcommand{\muv}{\boldsymbol{\mu}}
\newcommand{\const}{\mathrm{const}}
\newcommand{\uv}{\vec{\boldsymbol{u}}}
\newcommand{\q}{\tilde{\boldsymbol{q}}}
\newcommand{\ut}{\tilde{\boldsymbol{u}}}

\renewcommand{\P}{\prodi}
\newcommand{\A}{\mathscr{A}}
\newcommand{\B}{\mathscr{B}}
\newcommand{\K}{\mathcal{K}}
\newcommand{\m}{\mathfrak{m}}
\newcommand{\phis}{\upvarphi}
\newcommand{\Cs}{\mathscr{C}}
\newcommand{\Cc}{\mathscrbf{C}}
\newcommand{\vO}{\boldsymbol{0}}

\newcommand{\ie}{\emph{i.e.}\xspace}
\newcommand{\eg}{\emph{e.g.}\xspace}

\newcommand{\etal}{\emph{et al.}\xspace}

\renewcommand{\sim}{\thicksim}
\renewcommand{\div}{\grad\scal}
\newcommand{\divv}{\gradv\scal}

\newcommand{\Mat}{\mathrm{Mat}\,}
\newcommand{\scal}{\boldsymbol{\cdot}}
\newcommand{\grad}{\boldsymbol{\nabla}}

\newcommand{\abs}[1]{\lvert\, #1\, \rvert}

\newcommand{\norm}[1]{\lVert\, #1\, \rVert}

\newcommand{\gradv}{\bar{\boldsymbol{\nabla}}}

\newcommand{\pd}[2]{\frac{\partial #1}{\partial\/ #2}}

\newcommand{\eqdef}{\mathop{\stackrel{\,\mathrm{def}}{:=}\,}}

\newcommand{\half}{{\textstyle{1\over2}}}

\usepackage{acronym}

\begin{document}

\title[\Title]{Serre-type equations in deep water}

\author[D.~Dutykh]{Denys Dutykh$^*$}
\address{LAMA, UMR 5127 CNRS, Universit\'e Savoie Mont Blanc, Campus Scientifique, 73376 Le Bourget-du-Lac Cedex, France}
\email{Denys.Dutykh@univ-savoie.fr}
\urladdr{http://www.denys-dutykh.com/}
\thanks{$^*$ Corresponding author}

\author[D. Clamond]{Didier Clamond}
\address{Universit\'e de Nice -- Sophia Antipolis, Laboratoire J. A. Dieudonn\'e, Parc Valrose, 06108 Nice cedex 2, France}
\email{diderc@unice.fr}
\urladdr{http://math.unice.fr/~didierc/}

\author[M.~Chhay]{Marx Chhay}
\address{LOCIE, UMR 5271 CNRS, Universit\'e Savoie Mont Blanc, Campus Scientifique, 73376 Le Bourget-du-Lac Cedex, France}
\email{Marx.Chhay@univ-savoie.fr}
\urladdr{http://marx.chhay.free.fr/}



\begin{titlepage}
\thispagestyle{empty} 
\noindent
{\Large Denys \textsc{Dutykh}}\\
{\it\textcolor{gray}{CNRS--LAMA, Universit\'e Savoie Mont Blanc, France}}
\\[0.02\textheight]
{\Large Didier \textsc{Clamond}}\\
{\it\textcolor{gray}{Universit\'e de Nice -- Sophia Antipolis, LJAD, France}}
\\[0.02\textheight]
{\Large Marx \textsc{Chhay}}\\
{\it\textcolor{gray}{Polytech Annecy--Chamb\'ery, LOCIE, France}}
\\[0.08\textheight]

\vspace*{2cm}

\colorbox{Lightblue}{
  \parbox[t]{1.0\textwidth}{
    \centering\huge\sc
    \vspace*{0.7cm}
    
    \textcolor{bluepigment}{Serre-type equations in deep water}
    
    \vspace*{0.7cm}
  }
}

\vfill 

\raggedleft     
{\large \plogo} 
\end{titlepage}


\newpage
\thispagestyle{empty} 
\par\vspace*{\fill}   
\begin{flushright} 
{\textcolor{denimblue}{\textsc{Last modified:}} \today}
\end{flushright}


\newpage
\maketitle
\thispagestyle{empty}


\begin{abstract}

This manuscript is devoted to the modelling of water waves in the deep water regime with some emphasis on the underlying variational structures. The present article should be considered as a review of some existing models and modelling approaches even if new results are presented as well. Namely, we derive the deep water analogue of the celebrated \textsc{Serre}--\textsc{Green}--\textsc{Naghdi} equations which have become the standard model in shallow water environments.  The relation to existing models is discussed. Moreover, the multi-symplectic structure of these equations is reported as well. The results of this work can be used to develop various types of robust structure-preserving variational integrators in deep water. The methodology of constructing approximate models presented in this study can be naturally extrapolated to other physical flow regimes as well.

\bigskip
\noindent \textbf{\keywordsname:} deep water approximation; Serre--Green--Naghdi equations; variational principle; free surface impermeability \\


\end{abstract}


\newpage
\tableofcontents
\thispagestyle{empty}


\newpage
\section{Introduction}

Water waves is a special case of mechanical waves propagating at the interface of water and air. They play a central r\^ole in the interactions taking place between the ocean and atmosphere \cite{Komen1996, Thorpe2005}. R.~\textsc{Feynman} described the complexity of water waves using the following words \cite{Feynman2005}:
\begin{quote}
  \it[\,\dots] the next waves of interest, that are easily seen by everyone and which are usually used as an example of waves in elementary courses, are water waves. As we shall soon see, they are the worst possible example, because they are in no respects like sound and light; they have all the complications that waves can have [\,\dots]
\end{quote}
The complete mathematical formulation describing the propagation of water waves is quite complex to deal with. It cannot be solved analytically (unless in some asymptotic sense) and even efficient numerical algorithms are developed since 1970's \cite{Chan1970a} and nowadays this problem is far from being fully understood. That is why the water wave theory has always been developing by constructing sophisticated approximations \cite{Craik2004}. Traditionally we assume that the fluid is homogeneous (\ie its density $\rho\ =\ \const$) and \emph{ideal} and the flow is \emph{incompressible}. Additionally it is quite common to assume that the flow is also \emph{irrotational} \cite{Stoker1957}, \ie the flow vorticity vanishes identically.

\begin{figure}
  \centering
  \includegraphics[width=0.99\textwidth]{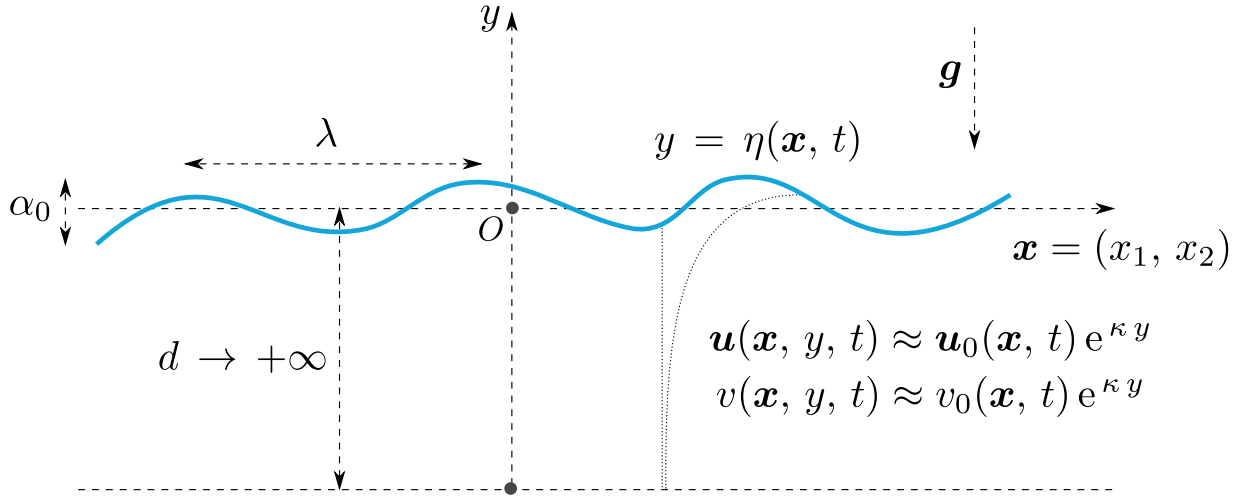}
  \caption{\small\em Sketch of the physical fluid domain.}
  \label{fig:sketch}
\end{figure}

In order to describe mathematically this problem let us introduce a \textsc{Cartesian} coordinate system $O\,x_1\,x_2\,y\,$, where the horizontal plane $O\,x_1\,x_2$ coincides with the still water level $y\ =\ 0\,$. The \emph{free surface} is given by the function $y\ =\ \eta\,(\x,\,t)\,$. The axis $O\,y$ points vertically upwards. The only force acting on the fluid is the gravity $\g\ =\ (0,\,0,\,-g)\,$. The sketch of the fluid domain is shown in Figure~\ref{fig:sketch}.

To have a compact notation we introduce the vector of horizontal independent variables $\x\ \eqdef\ (x_1,\,x_2)$ along with the associated horizontal gradient operator $\grad\ \eqdef\ \bigl(\partial_{\,x_1},\,\partial_{\,x_2}\bigr)\,$. The three-dimensional gradient will be denoted by $\gradv\ \eqdef\ \bigl(\partial_{\,x_1},\,\partial_{\,x_2},\,\partial_{\,y}\bigr)\,$. Similarly, the horizontal $\u\ =\ (u_1,\,u_2)$ and vertical $v$ components of the velocity can be separated for the sake of convenience. The three-dimensional velocity vector will be denoted by $\uv\ =\ (\u,\,v)\,$. Throughout this study two-dimensional vectors are denoted by bold symbols and three-dimensional vectors have additionally an over bar (or an arrow).

From the irrotationality assumption it follows that there exists a function $\phi(\x,\,y,\,t)$ called the velocity potential such that
\begin{equation*}
  \u\ =\ \grad\,\phi\,, \qquad v\ =\ \phi_{\,y}\,.
\end{equation*}
By taking into account also the flow incompressibility we obtain that the velocity potential $\phi$ is necessary a \emph{harmonic function}, \ie
\begin{equation}\label{eq:lapl}
  \grad^{\,2}\,\phi\ +\ \partial_{\,y}^{\,2}\,\phi\ =\ 0\,, \qquad
  (\x,\,y)\ \in\ \R^2\times\bigl(-\infty,\;\eta\,(\x,\,t)\bigr)\,.
\end{equation}
On the free surface we have the kinematic condition (also known as the free surface impermeability):
\begin{equation}\label{eq:kin}
  \eta_{\,t}\ +\ \grad\,\eta\scal\grad\,\phi\ =\ \partial_{\,y}\,\phi\,, \qquad
  y\ =\ \eta\,(\x,\,t)\,.
\end{equation}
The free surface being an \emph{isobar}, thus we have also the dynamic boundary condition:
\begin{equation}\label{eq:dyn}
  \phi_{\,t}\ +\ \half\,\abs{\grad\phi}^{\,2}\ +\ \half\,\phi_{\,y}^{\,2}\ +\ g\,\eta\ =\ 0\,, \qquad
  y\ =\ \eta\,(\x,\,t)\,.
\end{equation}
The last equation is known as the \textsc{Cauchy}--\textsc{Lagrange} integral in which we chose the gauge where the \textsc{Bernoulli} constant vanishes. 

Let us assume that the wave field consists mainly of waves with a characteristic wavelength $\lambda\,$, which corresponds to the characteristic wavenumber $\kappa\ \equiv\ \dfrac{2\,\pi}{\lambda}\,$. The average fluid depth is $d\,$. Thus, we can form a dimensionless number $\kappa\cdot d\,$. If this parameter $\kappa\cdot d\ \gg\ 1\,$ then we can further simplify our problem by applying the so-called \emph{deep water approximation} $d\ \to\ +\infty\,$, which allows to `evacuate' the solid impermeable bottom from the consideration. Basically, it allows to reduce the number of boundary conditions to satisfy by one. More precisely, we require that the fluid tends to the state of rest as we dive into it, \ie
\begin{equation*}
  \abs{\grad\,\phi}\ \to\ 0\,,\quad \phi_{\,y}\ \to\ 0\,, \quad \mbox{ as } \quad y\ \to\ -\infty\,.
\end{equation*}
In practice, even for values $\kappa\cdot d\ \apprge\ 2$ the deep water approximation can be already successfully applied in some situations \cite{Yuen1982}. In general, we refer to \cite{Yuen1982} as an excellent review on deep water waves. Equations \eqref{eq:lapl} -- \eqref{eq:dyn} will be referred to as the full water wave problem in deep water approximation.

The shallow water limit is much better understood nowadays. In shallow water there is a well-established hierarchy of hydrodynamic models:
\begin{itemize}
  \item nonlinear shallow water (\textsc{Saint}-\textsc{Venant} or \textsc{Airy} (especially in UK) or fully nonlinear non-dispersive) equations \cite{Airy1845, SV1871}
  \item \textsc{Boussinesq}-type (or weakly nonlinear weakly dispersive) equations \cite{Boussinesq1872}
  \item Fully nonlinear weakly dispersive equations \cite{Serre1953a}
  \item \dots
  \item Fully nonlinear fully dispersive \textsc{Euler} equations described above.
\end{itemize}
The deep water case is much less organized. The main difference between these two regimes comes from the dimensionless numbers which characterize the flow. When the depth $d$ is finite, we have one parameter $\dfrac{\alpha_0}{d}$ which characterizes the wave nonlinearity and another parameter $\dfrac{d^2}{\lambda^2}$ to describe flow `shallowness'. By applying asymptotic expansions in one or even two parameters we can obtain various approximate models. Now, if we take the limit $d\ \to\ +\infty$ both these parameters collapse in the deep water making this case somehow special. Traditionally, deep water waves have been described as perturbations of a certain carrier wave. In this way, the wave field has been conveniently described using wave envelopes \cite{Benney1967}. Then, the envelope function is shown to satisfy the nonlinear Schr\"odinger \cite{Zakharov1968} or \textsc{Dysthe}-type \cite{Dysthe1979} equations depending on the desired asymptotic order of accuracy. These equations can be also recast into the \textsc{Hamiltonian} framework \cite{Gramstad2011}. In the present study we take an alternative route without appealing to wave envelope techniques.

The present article is organized as follows. In Section~\ref{sec:var} we present three variational formulations of the full water wave problem \eqref{eq:lapl} -- \eqref{eq:dyn}. Then, in Section~\ref{sec:models} we briefly review the state of the art in deep water wave modelling and in Section~\ref{sec:deep} we present the variational derivations of \textsc{Saint}-\textsc{Venant} and \textsc{Serre} equations analogues in deep water. Finally, the main conclusions and perspectives of this study are discussed in Section~\ref{sec:disc}.


\section{Variational structures}
\label{sec:var}

In this Section we briefly describe the \emph{main} variational structures of the deep water wave problem in the chronological order of their appearance. Of course, this list is not being exhaustive.


\subsection{Lagrangian and Hamiltonian formulations}

In perfect agreement with the themes of this special issue, the water wave problem is known since A.~\textsc{Petrov} (1964) \cite{Petrov1964} and V.~\textsc{Zakharov} \cite{Zakharov1968} to have the \textsc{Hamiltonian} structure. Below we present the classical \textsc{Lagrangian} and \textsc{Hamiltonian} formulations together since they are naturally related by \textsc{Legendre} transformation. A more advanced \textsc{Luke}'s \textsc{Lagrangian} functional along with its generalizations will be presented below (see Sections \ref{sec:Luke} \& \ref{sec:relax}).

Let us compute the kinetic $\K$ and potential $\P$ energies of a deep fluid moving under the force of gravity $\g\,$:
\begin{equation*}
  \K\ \eqdef\ \frac{1}{2}\;\int_{\,\R^2}\int_{\,-\infty}^{\,\eta}\,\abs{\uv}^{\,2}\;\ud y\;\,\ud\x, \qquad
  \P\ =\ \frac{1}{2}\;g\,\int_{\,\R^2}\,\eta^{2}\;\ud\x\,.
\end{equation*}
According to \textsc{Hamilton}'s principle \cite{Basdevant2007}, the fluid motion has to provide a stationary value to the following action functional
\begin{equation}\label{eq:action}
  \S\ =\ \int_{\,t_0}^{\,t_1}\rho\,\L\;\ud t\,,
\end{equation}
where $\L$ is the \textsc{Lagrangian} density classically defined as
\begin{equation*}
  \L\ \eqdef\ \K\ -\ \P\,.
\end{equation*}
Below in Section~\ref{sec:Luke} we shall give another \textsc{Lagrangian} density. We have to keep in mind that the flow is incompressible, \ie
\begin{equation*}
  \divv\uv\ =\ 0\,,
\end{equation*}
and on the free surface we also have the kinematic boundary condition that we shall write as
\begin{equation*}
  \eta_{\,t}\ =\ \sqrt{1\ +\ \abs{\grad\eta}^{\,2}}\cdot u_{\,n}\,,
\end{equation*}
where $\u_n\ \eqdef\ \uv\scal\n$ is the normal velocity at the free surface and $\n$ is the outer unitary normal vector
\begin{equation*}
  \n\ =\ \frac{1}{\sqrt{1\ +\ \abs{\grad\eta}^{\,2}}}\;\begin{pmatrix}
    -\grad\eta \\
    1
  \end{pmatrix}\,.
\end{equation*}
We have to incorporate these conditions into \textsc{Hamilton} principle using two \textsc{Lagrange} multipliers $\phi\ =\ \phi(\x,\,y,\,t)$ and $\phis\ =\ \phis(\x,\,t)\,$:
\begin{equation*}
  \L\ =\ \K\ -\ \P\ +\ \int_{\,\R^2}\int_{\,-\infty}^{\,\eta}\phi\,\divv\uv\;\ud y\;\ud\x\ +\ \int_{\,\R^2}\bigl[\,\eta_{\,t}\ -\ \sqrt{1\ +\ \abs{\grad\eta}^{\,2}}\cdot u_{\,n}\,\bigr]\;\phis\;\ud\x\,.
\end{equation*}
By taking the variation of this functional with respect to $\uv$ and requiring that it vanishes in the fluid bulk we obtain
\begin{equation}\label{eq:duv}
  \delta\uv:\ \uv\ -\ \gradv\phi\ =\ \vO\,.
\end{equation}
Consequently, the flow is necessarily irrotational. It is a direct consequence of assumptions made above and the \textsc{Lagrange} multiplier $\phi$ is a velocity potential. From \textsc{Kelvin}'s circulation theorem we know that the flow initially irrotational will remain irrotational forever \cite{Batchelor1967}. The variational description of flows with vorticity is out of scope of the present study.

Taking into account \eqref{eq:duv}, from now on we can substitute $\uv\ =\ \gradv\phi$ into the \textsc{Lagrangian} density $\L\,$. By applying the \textsc{Gau\ss}--\textsc{Ostrogradsky} theorem to the \textsc{Lagrangian} density we obtain
\begin{equation*}
  \L\ =\ \int_{\,\R^2}\,\Bigl[\,\phis\,\eta_{\,t}\ +\ u_{\,n}\,\sqrt{1\ +\ \abs{\grad\eta}^{\,2}}\cdot\bigl(\phis\ -\ \phi\bigr\vert^{y\,=\,\eta}\bigr)\,\Bigr]\;\ud\x\ -\ \K\ -\ \P\,.
\end{equation*}
By taking the variation with respect to the normal velocity $u_{\,n}$ we obtain
\begin{equation*}
  \delta u_{\,n}:\ \phis\ -\ \phi\bigr\vert^{y\,=\,\eta}\ =\ 0\,.
\end{equation*}
Thus, the other \textsc{Lagrange} multiplier $\phis$ is simply the trace of the velocity potential at the free surface, \ie
\begin{equation*}
  \phis(\x,\,t)\ \equiv\ \phi\bigl(\x,\,y\,=\,\eta(\x,\,t),\,t\bigr)\,.
\end{equation*}
Finally, the \textsc{Lagrangian} density $\L$ becomes
\begin{equation*}
  \L\ =\ \int_{\,\R^2}\,\phis\,\eta_{\,t}\;\ud\x\ -\ \H\,
\end{equation*}
where $\H\ \eqdef\ \K\ +\ \P$ is the total fluid energy being also the \textsc{Hamiltonian} of the water wave problem:
\begin{equation*}
  \H\ =\ \frac{1}{2}\;\int_{\,\R^2}\,\int_{\,-\infty}^{\,\eta}\abs{\gradv\phi}^{\,2}\,\ud\x\ +\ \frac{1}{2}\;g\,\int_{\,\R^2}\,\eta^2\;\ud\x\,.
\end{equation*}
The last \textsc{Hamiltonian} functional was independently rediscovered by \textsc{Broer} (1974) \cite{Broer1974}, then by \textsc{Miles} (1977) \cite{Miles1977} and probably several other researchers.

The evolution equations for canonical variables are
\begin{equation*}
  \eta_{\,t}\ =\ \frac{\delta\,\H}{\delta\,\phis}\,, \qquad
  \phis_{\,t}\ =\ -\frac{\delta\,\H}{\delta\,\eta}\,.
\end{equation*}
By $\delta\,\H$ we denote the variational (\textsc{G\^ateaux}'s) derivative. In order to compute the \textsc{Hamiltonian} $\H$ one has to solve the \textsc{Laplace} equation
\begin{equation*}
  \gradv^{\,2}\phi\ =\ 0\,,
\end{equation*}
with corresponding boundary conditions:
\begin{equation*}
  \phi\bigr\vert^{y\,=\,\eta}\ =\ \phis\,, \qquad \abs{\gradv\phi}\ \to\ 0\,,\quad \mbox{ as }\quad y\ \to\ -\infty\,.
\end{equation*}
In general, it is not possible to solve this problem analytically. Consequently, in deep water one uses in practice asymptotic expansions with respect to the small parameter $\eps\ \sim\ \norm{\grad\eta}\ \sim\ \alpha_0\cdot\kappa\,$.


\subsection{Luke's Lagrangian formulation}
\label{sec:Luke}

In 1967 J.C.~\textsc{Luke} proposed to use the following functional \cite{Luke1967} (in finite depth case):
\begin{equation*}
  \L\ =\ \int_{\,-d}^{\,\eta}\bigl[\,\phi_{\,t}\ +\ \half\,\abs{\grad\phi}^{\,2}\ +\ \half\,\phi_{\,y}^{\,2}\ +\ g\,y\,\bigr]\;\ud y\,,
\end{equation*}
where $d$ is the constant water depth. The action integral is defined in \eqref{eq:action} as above. Without free surface effects this functional was proposed in 1929 by H.~\textsc{Bateman} \cite{Bateman1929}. One can easily recognize that the expression under the integral sign is the well-known \textsc{Cauchy}--\textsc{Lagrange} integral. In his seminal paper \cite{Luke1967} \textsc{Luke} justified the advantages of this functional over the classical \textsc{Lagrangian} $\L\ =\ \K\ -\ \P$ described above.

In order to apply the deep water approximation we have to take the limit $d\ \to\ +\infty\,$. The term $g\,y$ is not integrable, so before taking this limit we integrate it over the depth and remove the constant term $-g\;\dfrac{d^{\,2}}{2}$ which disappears under the \textsc{G\^ateaux} derivative operation. As a result, we obtain the following \textsc{Lagrangian} density:
\begin{equation}\label{eq:luke}
  \L\ =\ \int_{\,-\infty}^{\,\eta}\bigl[\,\phi_{\,t}\ +\ \half\,\abs{\grad\phi}^{\,2}\ +\ \half\,\phi_{\,y}^{\,2}\,\bigr]\;\ud y\;\ud\x\ +\ \frac{1}{2}\;g\,\eta^2\,.
\end{equation}
In order to recover the water wave problem equations \eqref{eq:lapl} -- \eqref{eq:dyn} in deep water, we write down the \textsc{Euler}--\textsc{Lagrange} equations corresponding to the functional \eqref{eq:luke}:
\begin{align*}
  \delta\phi:&\quad \grad^2\phi\ +\ \phi_{\,yy}\ =\ 0\,, \\
  \delta\phi\bigr\vert^{y\,=\,\eta}:&\quad \eta_{\,t}\ +\ \grad\phi\scal\grad\eta\ -\ \phi_{\,y}\ =\ 0\,, \\
  \delta\eta:&\quad \phi_{\,t}\ +\ \half\,\abs{\grad\phi}^{\,2}\ +\ \half\,\phi_{\,y}^{\,2}\ +\ g\,\eta\ =\ 0\,.
\end{align*}

\textsc{Luke}'s variational principle has at least one important advantage over the \textsc{Hamiltonian} principle: the flow incompressibility \eqref{eq:lapl} is incorporated into the variational principle and it does not have to be additionally assumed as a constraint. It appears as one of \textsc{Euler}--\textsc{Lagrange} equations.


\subsection{Relaxed Lagrangian formulation}
\label{sec:relax}

Recently, two authors of this manuscript proposed a generalization to \textsc{Luke}'s \textsc{Lagrangian} \cite{Clamond2009}. The so-called `\emph{relaxed variational principle}' will be extensively used in this study and we proceed to a brief description of the main ideas behind this generalization. Earlier in the literature this this method was introduced also under the name of a ``motivated \textsc{Legendre} transform'' (see \eg \cite{Henyey1983} for more details). In this Section we follow closely the broad lines of our previous publication \cite{Clamond2009}.

We would like to introduce more variables into the \textsc{Luke} \textsc{Lagrangian} \eqref{eq:luke} which has the velocity potential $\phi(\x,\,y,\,t)$ and free surface elevation $\eta(\x,\,t)$ in its original form. Let us introduce also explicitly the components of the velocity field $\u\ =\ \grad\phi$ and $v\ =\ \phi_{\,y}$ by using two \textsc{Lagrange} multipliers $\muv$ and $\nu\,$:
\begin{equation*}
  \L\ =\ -\phis\,\eta_{\,t}\ +\ \frac{1}{2}\;g\,\eta^{2}\ +\ \int_{\,-\infty}^{\,\eta}\,\Bigl[\,\half\,(\,\u^2\ +\ v^2\,)\ +\ \mu\scal(\grad\phi\ -\ \u)\ +\ \nu\,(\phi_{\,y}\ -\ v)\,\Bigr]\;\ud y\,,
\end{equation*}
where we took also the term $\phi_{\,t}$ out of the integral sign for the sake of convenience. By applying the \textsc{Gau\ss}--\textsc{Ostrogradsky} theorem we can rewrite the \textsc{Lagrangian} $\L$ in the following \emph{equivalent} form:
\begin{align*}
  \L\ &=\ -\bigl(\eta_{\,t}\ +\ \tilde{\muv}\scal\grad\eta\ -\ \tilde{\nu}\bigr)\,\phis\ +\ \frac{1}{2}\;g\,\eta^2\ + \\
  & \quad \int_{\,-\infty}^{\,\eta}\,\Bigl[\,\half\,(\u^2\ +\ v^2)\ -\ \muv\scal\u\ -\ \nu\cdot v\ -\ (\div\muv\ +\ \nu_{\,y})\,\phi\,\Bigr]\;\ud y\,.
\end{align*}
The tildes denote the quantities evaluated at the free surface, \ie $\tilde{\nu}(\x,\,t)\ \eqdef\ \nu\bigl(\x,\,y\,=\,\eta(\x,\,t),\,t\bigr)\,$. The last functional $\L$ is the so-called \emph{relaxed variational principle}. Let us count the degrees of freedom:
\begin{enumerate}
 \item $\eta\,(\x,\,t)$ is the free surface elevation
 \item $\phi(\x,\,y,\,t)$ is the velocity potential
 \item $\u(\x,\,y,\,t)$ is the horizontal velocities vector
 \item $v(\x,\,y,\,t)$ is the vertical velocity
 \item $\muv(\x,\,y,\,t)$ is the \textsc{Lagrange} multiplier associated to the horizontal velocities
 \item $\nu(\x,\,y,\,t)$ is the \textsc{Lagrange} multiplier associated to the vertical velocity
\end{enumerate}
So, instead of having two degrees of freedom in the original \textsc{Luke} \textsc{Lagrangian}, the relaxed \textsc{Lagrangian} has six. This extra freedom can be used to derive various approximations which was illustrated in \cite{Clamond2009}.

\subsubsection{Lagrange multipliers}

For the purposes of the present study we may content with four degrees of freedom by eliminating the \textsc{Lagrange} multipliers $\muv$ and $\nu$. Indeed, let us compute the variations of the relaxed \textsc{Lagrangian} with respect to $\u$ and $v$:
\begin{align*}
  \delta\u:& \quad \u\ -\ \muv\ =\ \vO\,, \\
  \delta v:& \quad v\ -\ \nu\ =\ 0\,.
\end{align*}
This computation gives us also the physical sense of \textsc{Lagrange} multipliers --- they are pseudo-velocities, which coincide with physical velocities $\u$ and $v$ at least in the \emph{unconstrained} case. Thus, we can substitute $\muv\ =\ \u$ and $\nu\ =\ v$ into $\L$ to obtain
\begin{equation}\label{eq:relax}
  \L\ =\ -\bigl(\eta_{\,t}\ +\ \tilde{\u}\scal\grad\eta\ -\ \tilde{v}\bigr)\,\phis\ +\ \frac{1}{2}\;g\,\eta^2\ -\ \int_{\,-\infty}^{\,\eta}\,\Bigl[\,\half\,(\u^2\ +\ v^2)\ -\ (\div\u\ +\ v_{\,y})\,\phi\,\Bigr]\;\ud y\,.
\end{equation}
We shall use extensively this \textsc{Lagrangian} in the rest of this manuscript.


\bigskip
\paragraph*{Intermediate conclusions.}

The variational structure in general (such as \textsc{Hamiltonian} or \textsc{Lagrangian} functionals) is important in many respects. First of all, since the full equations \eqref{eq:lapl} -- \eqref{eq:dyn} enjoy this variational structure, we should seek for approximate models which enjoy the same structure and, thus, preserve some sub-set of qualitative properties of the base model. For instance, the \textsc{Hamiltonian} formalism \cite{Zakharov1997} allows to simplify asymptotic developments in powers of the nonlinearity parameter $\eps\ \eqdef\ \dfrac{\alpha_0}{\lambda}\,$, which is the wave steepness in the deep water regime. Finally, the \textsc{Hamiltonian} formulation allows also to put the problem of hydrodynamic waves in a unified framework of nonlinear waves in various media \cite{Zakharov1997, Zakharov2009a}. Thus, methods developed in other fields might be directly transposed to water waves.


\section{State of the art}
\label{sec:models}

The golden standard in deep water wave modelling is incontestably the cubic \textsc{Zakharov} model used, recently \eg in \cite{Dyachenko2003, Korotkevich2008} to study wave (weak) turbulence \cite{Zakharov1992}. These equations are obtained by expanding the \textsc{Hamiltonian} in the wave steepness parameter
\begin{equation*}
  \H\ =\ \H_{\,0}\ +\ \H_{\,1}\ +\ \H_{\,2}\ +\ \ldots
\end{equation*}
The cubic \textsc{Zakharov} equations are obtained by truncating this expansion after quartic terms (thus, the governing equations are effectively cubic after taking the variations). This model is weakly nonlinear but it is valid for the whole spectrum of gravity waves. Cubic \textsc{Zakharov} equations are well understood nowadays. Consequently, in the present study we focus on models which do the opposite: on one hand, there are \emph{a priori} no assumptions on the nonlinearity parameter, on the other hand, we describe waves around certain wavenumber $\kappa\,$. Let us review the state of the art by following the main steps of \cite{Kraenkel2005}. However, below we generalize their computations to the three-dimensional case.

Consider the incompressible \textsc{Euler} equations:
\begin{align}\label{eq:inc}
  \div\u\ +\ v_{\,y}\ &=\ 0\,, \\
  \dot{\u}\ +\ \grad p\ &=\ \vO\,,\label{eq:hor} \\
  \dot{v}\ +\ p_{\,y}\ +\ g\ &=\ 0\,,\label{eq:vert}
\end{align}
where $p$ is the fluid pressure and the over dot denotes the total material derivative, \ie
\begin{equation*}
  \dot{(\cdot)}\ \eqdef\ (\cdot)_{\,t}\ +\ \u\scal\grad(\cdot)\ +\ v\,(\cdot)_{\,y}\,.
\end{equation*}
The governing equations are completed with the following boundary conditions:
\begin{align}\label{eq:kin}
  \eta_{\,t}\ +\ \u\scal\grad\eta\ &=\ v\,, \qquad y\ =\ \eta\,(\x,\,t)\,, \\
  p\ &=\ p_a\,, \qquad y\ =\ \eta\,(\x,\,t)\,,\label{eq:dyn} \\
  \abs{\u},\, \abs{v}\ &\to\ 0\,, \qquad y\ \to\ -\infty\,,
\end{align}
where $p_a$ is the constant atmospheric pressure.

\begin{figure}
  \centering
  \subfigure[$u_{\,1}\,(x_1,\,y,\,t)$]{\includegraphics[width=0.49\textwidth]{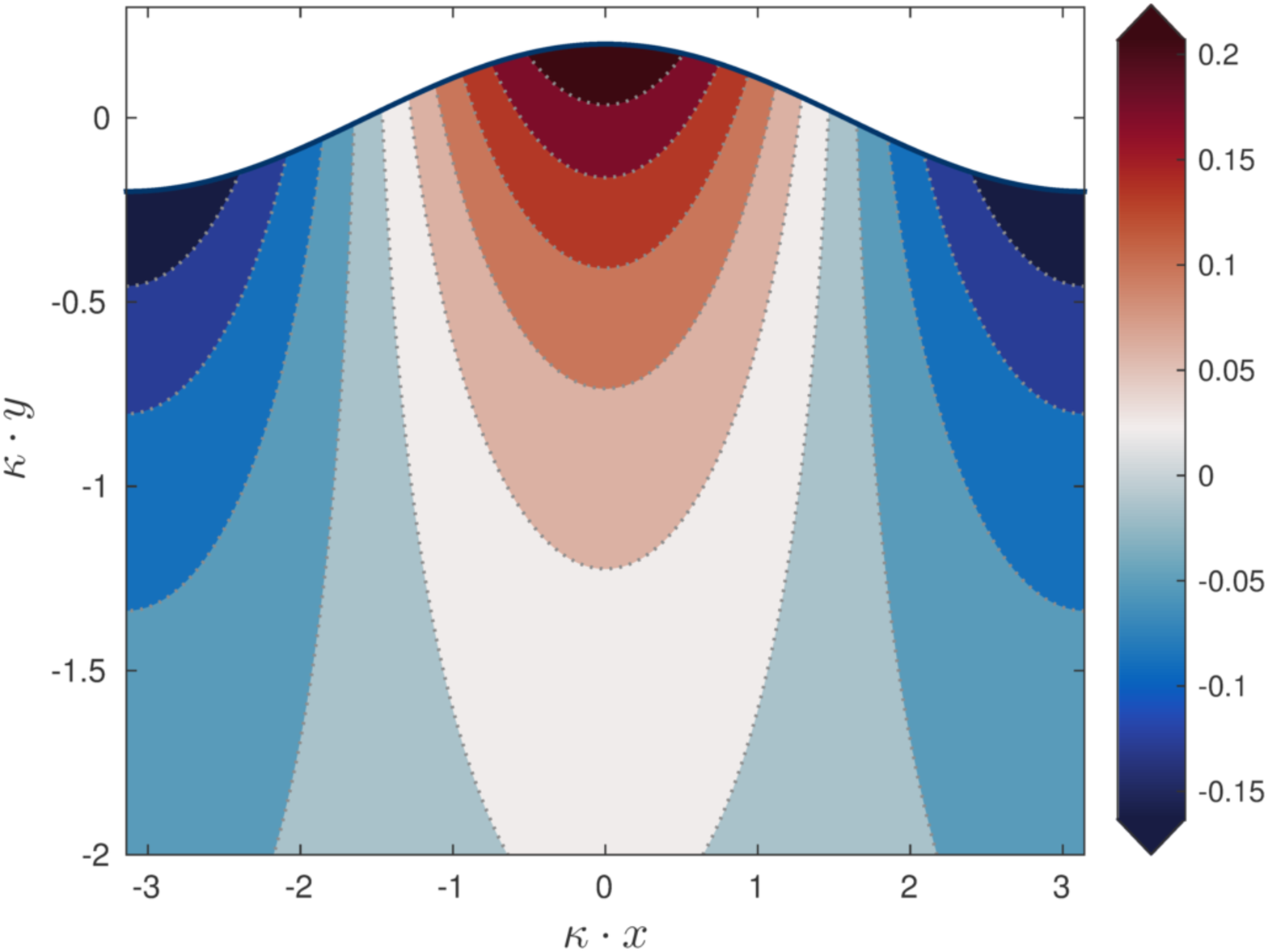}}
  \subfigure[$v\,(x_1,\,y,\,t)$]{\includegraphics[width=0.49\textwidth]{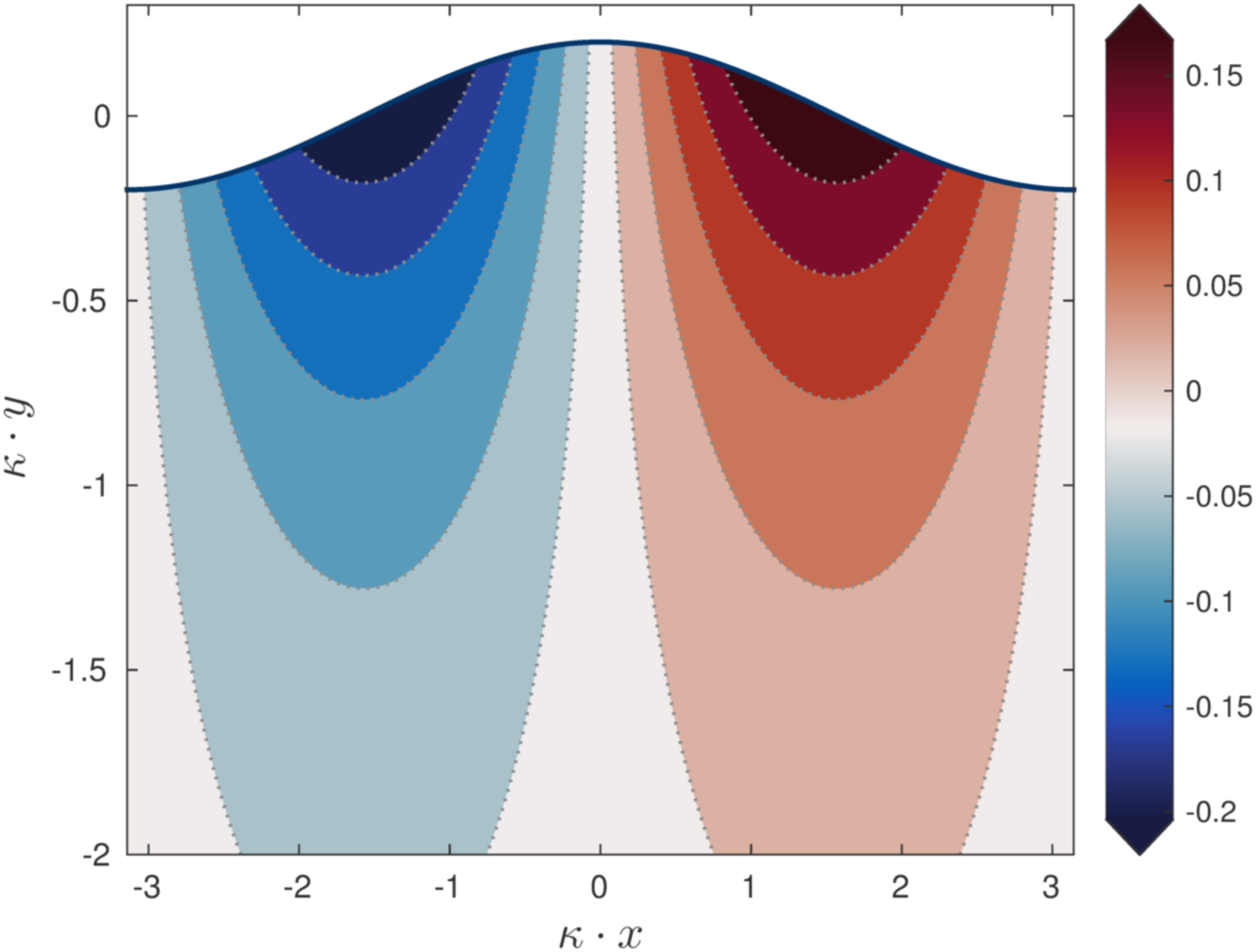}}
  \caption{\small\em Vertical structure of the chosen ansatz \eqref{eq:ans} under a simple linear travelling periodic wave: (a) the horizontal and (b) vertical components of the velocity field. The periodic wave amplitude is $\alpha\ =\ 0.2\,$.}
  \label{fig:ans}
\end{figure}

In order to derive an approximate model, \textsc{Kraenkel} \etal \cite{Kraenkel2005} propose to take the following solution ansatz for the \textsc{Euler} equations \eqref{eq:inc} -- \eqref{eq:vert}:
\begin{equation}\label{eq:ans}
  \u(\x,\,y,\,t)\ =\ \u_{\,0}(\x,\,t)\;\ue^{\,\kappa\,y}\,, \quad v(\x,\,y,\,t)\ =\ -\frac{1}{\kappa}\;(\div\u_{\,0})\;\ue^{\,\kappa\,y}\,.
\end{equation}
From above ansatz it is straightforward to understand the physical sense of the variable $\u_0$ --- it is simply the value of the 3D horizontal velocity $\u$ on the surface $y\ =\ 0\,$. Other ans\"atze will be considered below. Here, $\kappa\ =\ \const$ is the wavenumber around which we model water waves in the spectral domain. The vertical velocity ansatz is chosen to satisfy identically the free surface incompressibility \eqref{eq:inc}. The velocity fields under a linear periodic wave predicted by this ansatz is represented in Figure~\ref{fig:ans}.

\begin{figure}
  \centering
  \subfigure[$(p\,-\,p_a)\,(x_1,\,y,\,t)$]{\includegraphics[width=0.49\textwidth]{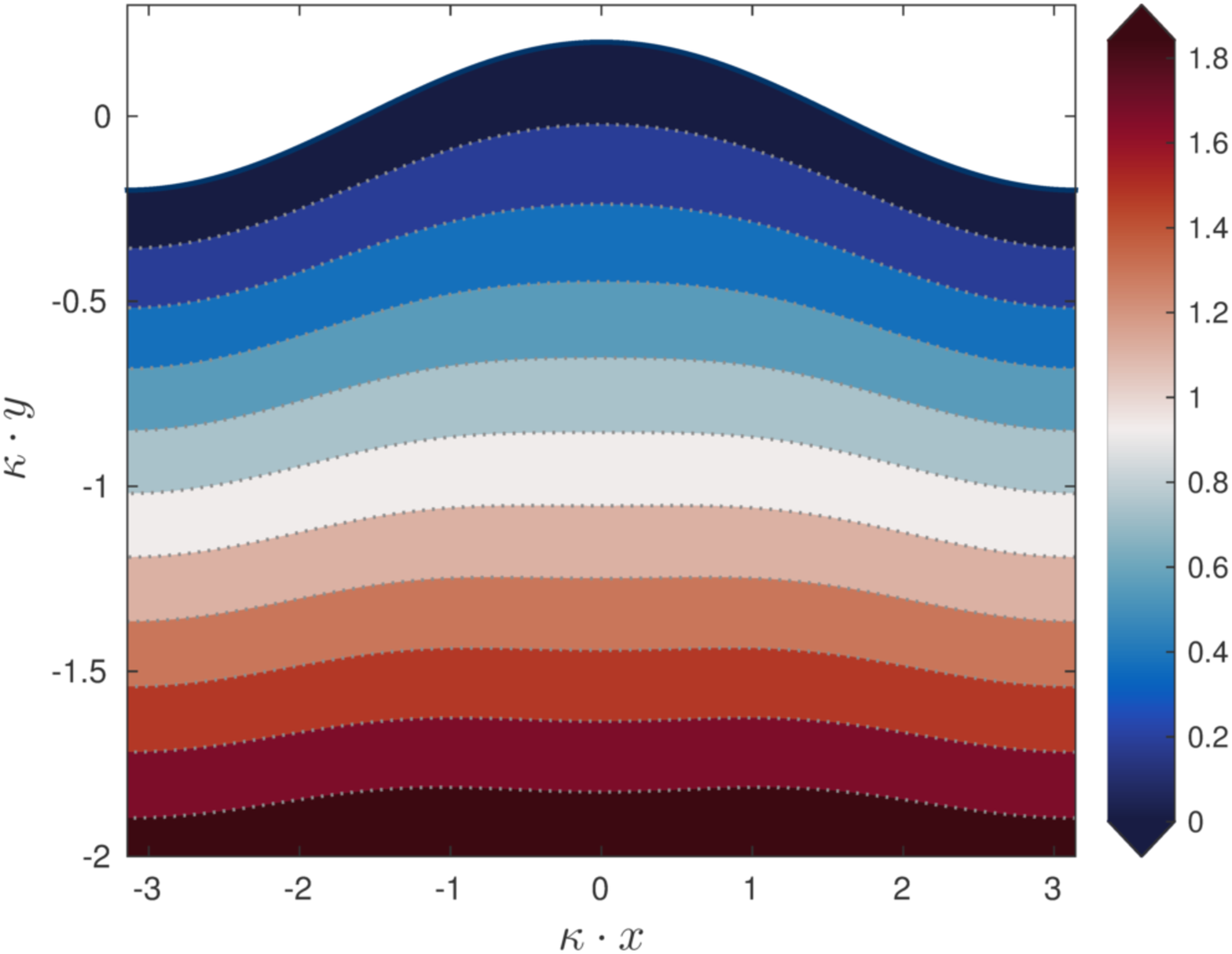}}
  \subfigure[$p_d\,(x_1,\,y,\,t)$]{\includegraphics[width=0.49\textwidth]{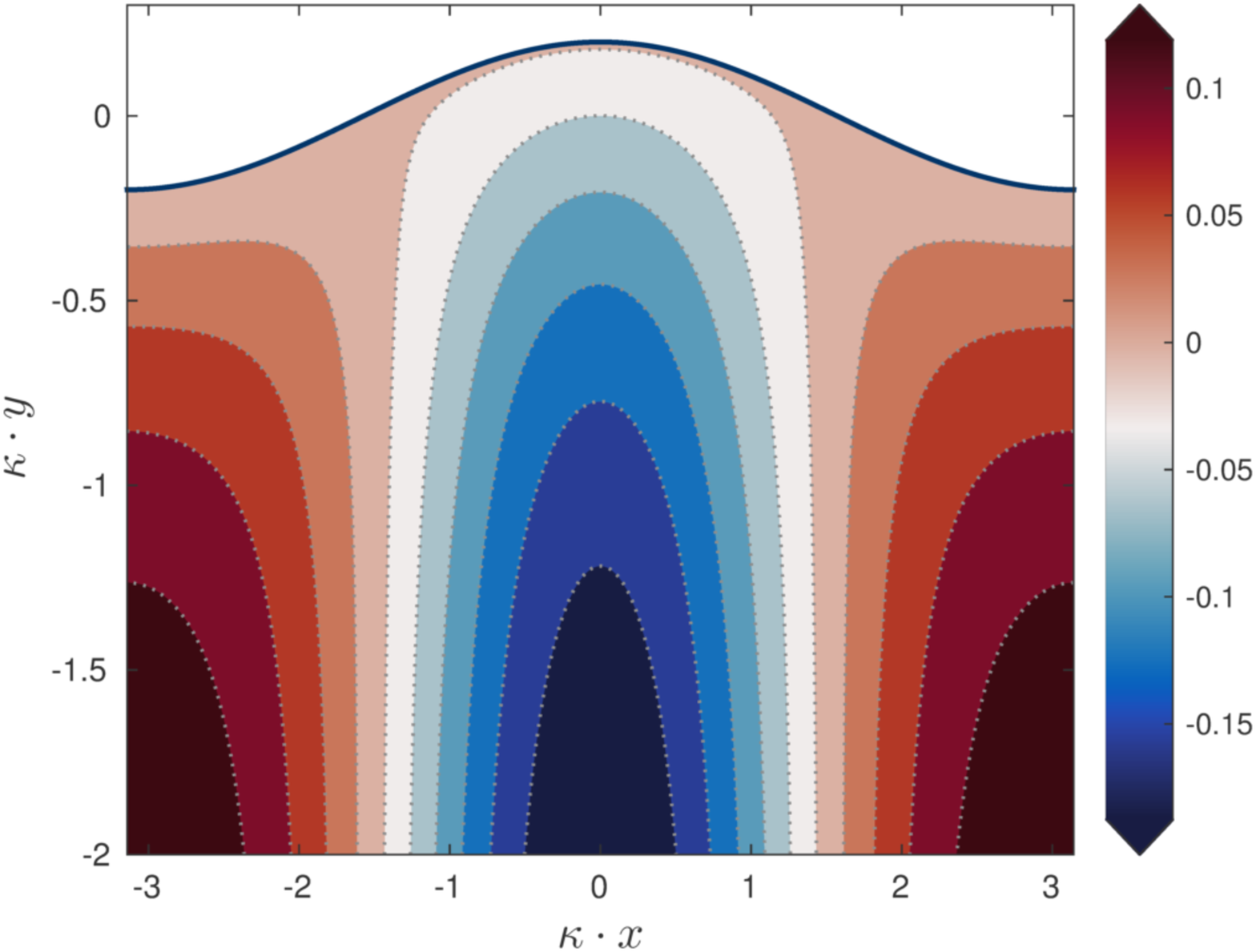}}
  \caption{\small\em Fluid pressure distribution under a linear travelling wave as predicted by equation \eqref{eq:press} (a). The right panel (b) shows the dynamic pressure $p_d\ \eqdef\ p\ -\ p_a\ -\ g\,(\eta\ -\ y)$ distribution (\ie without hydrostatic effects). The periodic wave amplitude is $\alpha\ =\ 0.2\,$.}
  \label{fig:press}
\end{figure}

Substituting ansatz \eqref{eq:ans} into the kinematic boundary condition \eqref{eq:kin} we readily obtain the mass conservation equation:
\begin{equation}\label{eq:mass}
  \kappa\,\eta_{\,t}\ +\ \div\bigl[\,\u_{\,0}\,\ue^{\,\kappa\,\eta}\,\bigr]\ =\ 0\,.
\end{equation}
In order to derive momentum balance equations, we compute first the material derivatives using ansatz \eqref{eq:ans}:
\begin{align*}
  \dot{\u}\ &=\ \u_{\,0\,t}\;\ue^{\,\kappa\,y}\ +\ \Cc\,\ue^{\,2\,\kappa\,y}\,,\\
  -\kappa\,\dot{v}\ &=\ \A\,\ue^{\,\kappa\,y}\ +\ \B\,\ue^{\,2\,\kappa\,y}\,,
\end{align*}
where $\A\,$, $\B$ and $\Cc\ =\ \bigl(\Cs_{\,1},\,\Cs_{\,2}\bigr)$ are defined as
\begin{equation*}
  \A\ =\ \div\u_{\,0\,t}\,, \qquad
  \B\ =\ \grad(\div\u_{\,0})\scal\u_{\,0}\ -\ (\div\u_{\,0})^{\,2}\,,
\end{equation*}
\begin{equation*}
  \Cs_{\,1}\ =\ u_{\,2}\;\pd{u_{\,1}}{x_2}\ -\ u_{\,1}\;\pd{u_2}{x_2}\,, \qquad
  \Cs_{\,2}\ =\ u_{\,1}\;\pd{u_{\,2}}{x_1}\ -\ u_{\,2}\;\pd{u_1}{x_1}\,.
\end{equation*}
By substituting $\dot{v}$ into \eqref{eq:vert} and taking into account the boundary condition \eqref{eq:dyn}, we obtain the pressure distribution in the fluid bulk:
\begin{equation}\label{eq:press}
  p\ -\ p_a\ =\ g\,(\eta\ -\ y)\ +\ \frac{1}{\kappa^2}\;\Bigl[\,\bigl(\ue^{\,\kappa\,y}\ -\ \ue^{\,\kappa\,\eta}\bigr)\;\A\ +\ \frac{1}{2}\,\bigl(\ue^{\,2\,\kappa\,y}\ -\ \ue^{\,2\,\kappa\,\eta}\bigr)\;\B\,\Bigr]\,.
\end{equation}
Notice that the pressure field $p$ diverges when $y\ \to\ -\infty$ due to the hydrostatic effects (in agreement with the \textsc{Archimedes} law). The pressure distribution under a linear periodic wave is shown in Figure~\ref{fig:press}.

The problem now is to satisfy in some sense equation \eqref{eq:hor} with available expressions for $\dot{\u}$ and $p\,$. \textsc{Kraenkel} \etal proposed the following weak formulation:
\begin{equation*}
  \int_{\,-\infty}^{\,\eta}\Bigl[\,\dot{\u}\ +\ \grad p\,\Bigr]\;\ue^{\,\m\,y}\;\ud y\ =\ \vO\,,
\end{equation*}
where $\m\ >\ 0$ is a modelling parameter to be chosen later. So, the \textsc{Newton} law is satisfied in an average sense. The exponential weight function allows to overcome the problem of \textsc{Archimedean} pressure divergence. In order to derive the momentum equation in a conservative form, we shall use the equivalent form:
\begin{equation*}
  \int_{\,-\infty}^{\,\eta}\,\dot{\u}\;\ue^{\,\m\,y}\;\ud y\ +\ \grad\,\biggl[\;\int_{\,-\infty}^{\,\eta}\,p\;\ue^{\,\m\,y}\;\ud y\;\biggr]\ -\ p_a\,\ue^{\,\m\,\eta}\cdot\grad\eta\ =\ \vO\,.
\end{equation*}
After performing all computations we obtain the desired horizontal momentum equation:
\begin{multline}\label{eq:moment}
  \frac{\ue^{\,(\m\, +\, \kappa)\,\eta}}{\m\, +\, \kappa}\;\Bigl[\,\u_{\,0\,t}\ +\ \frac{\m\ +\ \kappa}{\m\ +\ 2\,\kappa}\;\ue^{\,\kappa\,\eta}\;\Cc\,\Bigr]\ +\\ \grad\,\biggl[\,g\;\frac{\ue^{\,\m\,\eta}}{\m^2}\ -\ \A\;\frac{\ue^{\,(\m\,+\,\kappa)\,\eta}}{\kappa\,\m\,(\m\ +\ \kappa)}\ -\ \frac{1}{2}\;\B\;\frac{\ue^{\,(\m\, +\, 2\,\kappa)\,\eta}}{\kappa\,\m\,(\m\ +\ 2\,\kappa)}\,\biggr]\ =\ \vO\,.
\end{multline}
Just derived equations \eqref{eq:mass} and \eqref{eq:moment} constitute a closed system which describes the evolution of water waves in deep water.

\begin{remark}
In order to make a check of the derivation made above, we shall restrict our attention to the two-dimensional case. Here we can set $u_{\,0\,1}\ \rightsquigarrow\ u\,$, $u_{\,0\,2}\ \rightsquigarrow\ 0$ and dependent coefficients become:
\begin{equation*}
  \A\ \rightsquigarrow\ u_{\,x\,t}\,, \qquad
  \B\ \rightsquigarrow\ u\,u_{\,x\,x}\ -\ u_{\,x}^{\,2}\,, \qquad
  \Cc\ \rightsquigarrow\ \vO\,.
\end{equation*}
Thus, equations \eqref{eq:mass} and \eqref{eq:moment} in 2D become:
\begin{align}\label{eq:mass1}
  \kappa\,\eta_{\,t}\ +\ \bigl[\,u\,\ue^{\,\kappa\,\eta}\,\bigr]_{\,x}\ &=\ 0\,, \\
  \frac{\ue^{\,(\m\, +\, \kappa)\,\eta}}{\m\, +\, \kappa}\;u_{\,t}\ +\ \biggl[\,g\;\frac{\ue^{\,\m\,\eta}}{\m^2}\ -\ u_{\,x\,t}\;\frac{\ue^{\,(\m\,+\,\kappa)\,\eta}}{\kappa\,\m\,(\m\, +\, \kappa)}\ -\ \bigl(u\,u_{\,x\,x}\ -\ u_{\,x}^{\,2}\bigr)\;\frac{\ue^{\,(\m\,+\,2\,\kappa)\,\eta}}{\kappa\,\m\,(\m\, +\, 2\,\kappa)}\,\biggr]_{\,x}\ &=\ 0\,.\label{eq:moment1}
\end{align}
By switching to dimensionless variables ($g\ \rightsquigarrow\ 1\,$, $\kappa\ \rightsquigarrow\ 1$) we recover equations (2.18) and (2.20) from \cite{Kraenkel2005}.
\end{remark}

As we saw above, the water wave problem possesses \textsc{Hamiltonian} and \textsc{Lagrangian} variational structures. The question we can ask is whether just derived equations \eqref{eq:mass}, \eqref{eq:moment} (which are supposed to approximate the full \textsc{Euler} equations) possess at least one of these structures? By looking at equations \eqref{eq:mass}, \eqref{eq:moment} the answer is not clear. Even in `Conclusion and comments' Section in \cite{Kraenkel2005} the Authors admit that they did not succeed in finding a \textsc{Hamiltonian} formulation even for the short wave limit of these equations. We shall propose some fixes to this problem below in Section~\ref{sec:varder}.


\subsection{Choice of the modelling parameter}

In order to derive a physically sound value for the modelling parameter $\m$, we consider the governing equations in 2D and we linearize equations \eqref{eq:mass1}, \eqref{eq:moment1}:
\begin{align*}
  \kappa\,\eta_{\,t}\ +\ u_{\,x}\ &=\ 0\,, \\
  \frac{1}{\m\,+\,\kappa}\;u_{\,t}\ +\ \biggl[\,\frac{g}{\m}\;\eta\ -\ \frac{1}{\kappa\,\m\,(\m\,+\,\kappa)}\;u_{\,x\,t}\,\biggr]_{\,x}\ &=\ 0\,.
\end{align*}
It is easy to eliminate the variable $\eta$ from the above equations to obtain the linear version of the so-called \emph{improved} \textsc{Boussinesq} \emph{equation}:
\begin{equation*}
  \m\,\kappa\,u_{\,t\,t}\ -\ g\,(\m\ +\ \kappa)\,u_{\,x\,x}\ -\ u_{\,t\,t\,x\,x}\ =\ 0\,.
\end{equation*}
Then, we look for plane wave solutions of the form:
\begin{equation*}
  u(x,\,t)\ =\ \alpha_0\,\ue^{\,\ui\,(k\,x\ -\ \omega\,t)}\,.
\end{equation*}
Such solutions exist only if wave frequency $\omega$ and wavenumber $k$ are related by the following relation, which is called the \emph{dispersion relation}:
\begin{equation*}
  c_p(k)\ =\ \frac{\omega(k)}{k}\ =\ \sqrt{g\;\frac{\m\ +\ \kappa}{\m\,\kappa\ +\ k^{\,2}}}\,.
\end{equation*}
The ratio $c_p(k)$ is called the \emph{phase velocity}. Notice that in the limit $k\ \to\ \kappa$ we obtain
\begin{equation*}
  \lim_{k\, \to\, \kappa}\frac{\omega(k)}{k}\ =\ \sqrt{\frac{g}{\kappa}}\,,
\end{equation*}
which coincides with the exact phase velocity of the full \textsc{Euler} equations in deep water. Finally, in order to determine the modelling parameter $\m$ we consider the \emph{group velocity}:
\begin{equation*}
  c_g(k)\ \eqdef\ \pd{\omega(k)}{k}\,.
\end{equation*}
In the limit $k\ \to\ \kappa$ we obtain
\begin{equation*}
  \lim_{k\, \to\, \kappa} c_g(k)\ =\ \frac{\m}{\m\ +\ \kappa}\cdot\sqrt{\frac{g}{\kappa}}\,.
\end{equation*}
Now it is straightforward to notice that we recover the exact expression $c_g(\kappa)\ =\ \dfrac{1}{2}\cdot\sqrt{\dfrac{g}{\kappa}}$ for deep water \textsc{Euler} equations only if $\m\ \equiv\ \kappa\,$. This is the desired value of the modelling parameter $\m\,$.


\subsection{Variational derivations}
\label{sec:varder}

In this Section we propose alternative derivations of the model equations similar to \eqref{eq:mass}, \eqref{eq:moment}, but based on the relaxed variational principle \eqref{eq:relax}. Of course, the resulting models obtained through variational derivation might (and actually will) be different from \eqref{eq:mass}, \eqref{eq:moment}. However, the advantage here is that the \textsc{Lagrangian} structure is preserved by construction.

\subsubsection{Weakly compressible ansatz}
\label{sec:weak}

Consider an ansatz similar to \eqref{eq:ans}:
\begin{equation}\label{eq:ans1}
  \phi(\x,\,y,\,t)\, =\, \phi_{\,0}(\x,\,t)\;\ue^{\,\kappa\,y}\,, \ \u(\x,\,y,\,t)\, =\, \u_{\,0}(\x,\,t)\;\ue^{\,\kappa\,y}\,,\ v(\x,\,y,\,t)\, =\, v_{\,0}(\x,\,t)\;\ue^{\,\kappa\,y}\,.
\end{equation}
The main difference with \eqref{eq:ans} is that here the vertical velocity approximation $v_0$ is kept independent from $\u_0$ and it will be chosen by the variational principle. Substituting \eqref{eq:ans1} into \textsc{Lagrangian} \eqref{eq:relax} and performing exactly all the integrations over the depth, we obtain the following \textsc{Lagrangian} density:
\begin{align}\label{eq:incL}
  \L\ &=\ -\Bigl[\,\eta_{\,t}\ +\ \bigl(\u_{\,0}\scal\grad\eta\ -\ v_{\,0}\bigr)\;\ue^{\,\kappa\,\eta}\,\Bigr]\;\phi_0\,\ue^{\,\kappa\,\eta}\ +\ \frac{1}{2}\;g\,\eta^2\nonumber\\
  &\qquad -\ \frac{1}{2\,\kappa}\;\Bigl[\,\half\,(\abs{\u_{\,0}}^{\,2}\ +\ v_{\,0}^{\,2})\ -\ (\div\u_{\,0}\ +\ \kappa\,v_{\,0})\,\phi_{\,0}\,\Bigr]\;\ue^{\,2\,\kappa\,\eta}\,.
\end{align}
The governing equations are then obtained by computing variations of this functional:
\begin{align*}
  \delta\phi_0:& \quad \eta_{\,t}\ +\ \bigl(\u_{\,0}\scal\grad\eta\ -\ v_{\,0}\bigr)\;\ue^{\,\kappa\,\eta}\ =\ \frac{1}{2\,\kappa}\;\bigl(\div\u_{\,0}\ +\ \kappa\,v_{\,0}\bigr)\;\ue^{\,\kappa\,\eta}\,, \\
  \delta\u_{\,0}:& \quad \u_{\,0}\ +\ \grad\phi_{\,0}\ +\ 4\,\kappa\,\phi_{\,0}\,\grad\eta\ =\ \vO\,, \\
  \delta v_{\,0}:& \quad \phi_{\,0}\ -\ \frac{1}{2\,\kappa}\;\bigl(v_{\,0}\ -\ \kappa\,\phi_{\,0}\bigr)\ =\ 0\,, \\
  \delta\eta:& \quad \bigl(\phi_{\,0}\,\ue^{\,\kappa\,\eta}\bigr)_{\,t}\ +\ \div\bigl[\,\phi_{\,0}\,\u_{\,0}\,\ue^{2\,\kappa\,\eta}\,\bigr]\ +\ g\,\eta\\
  &\quad -\ \Bigl[\,\half\,(\abs{\u_{\,0}}^{\,2}\ +\ v_{\,0}^{\,2})\ -\ (\div\u_{\,0}\ +\ \kappa\,v_{\,0})\,\phi_{\,0}\,\Bigr]\;\ue^{\,2\,\kappa\,\eta} \\
  &\quad -\ \kappa\,\Bigl[\,\eta_{\,t}\ +\ 2\,\bigl(\u_{\,0}\scal\grad\eta\ -\ v_{\,0}\bigr)\;\ue^{\,\kappa\,\eta}\,\Bigr]\,\phi_{\,0}\,\ue^{\,\kappa\,\eta}\ =\ 0\,.
\end{align*}
The \textsc{Euler}--\textsc{Lagrange} equations turn out to be rather complicated. However, we can learn some lessons nevertheless. The variation $\delta\u_{\,0}$ gives us the connection between the horizontal velocity $\u_{\,0}$ and the velocity potential $\phi_{\,0}$ (thus, $\u_{\,0}$ can be in principle eliminated from the equations). In particular, one can see that the flow is \emph{not} irrotational. The variation $\delta v_{\,0}$ gives us the expression of the vertical velocity $v_{\,0}\ =\ 3\,\kappa\,\phi_{\,0}$ in terms of the velocity potential. The variation with respect to $\eta$ gives the analogue of the \textsc{Cauchy}--\textsc{Lagrange} integral (\ie an \emph{unsteady} \textsc{Bernoulli} equation). Finally, the variation with respect to $\phi_{\,0}$ gives us the mass conservation equation and in order to have a conservative form, it is better to reinforce the flow incompressibility, \ie
\begin{equation*}
  \div\u_{\,0}\ +\ \kappa\,v_{\,0}\ \equiv\ 0\,.
\end{equation*}
It will be done in the following Section.


\subsubsection{Exactly incompressible ansatz}

Now we take the same ansatz \eqref{eq:ans1}, but the vertical velocity approximation $v_{\,0}$ is chosen in order to satisfy identically the incompressibility condition. It is not difficult to see that this goal can be achieved by taking $v_{\,0}\ \equiv\ -\frac{1}{\kappa}\;\div\u_{\,0}\,$. In this way we recover ansatz \eqref{eq:ans} proposed by \textsc{Kraenkel} \etal \cite{Kraenkel2005}. Substituting this expression for $v_{\,0}$ into the \textsc{Lagrangian} density \eqref{eq:incL}, we obtain the following slightly more compact density functional:
\begin{align*}
  \L\ &=\ -\Bigl[\,\eta_{\,t}\ +\ \bigl(\u_{\,0}\scal\grad\eta\ +\ \frac{1}{\kappa}\;\div\u_{\,0}\bigr)\;\ue^{\,\kappa\,\eta}\,\Bigr]\;\phi_0\,\ue^{\,\kappa\,\eta}\ +\ \frac{1}{2}\;g\,\eta^2\\
  &\qquad -\ \frac{1}{4\,\kappa}\;\Bigl[\,\abs{\u_{\,0}}^{\,2}\ +\ \frac{1}{\kappa^2}\,(\div\u_{\,0})^{\,2}\,\Bigr]\;\ue^{\,2\,\kappa\,\eta}\,.
\end{align*}
The \textsc{Euler}--\textsc{Lagrange} equations yield the following system:
\begin{align*}
  \delta\phi_{\,0}:&\quad \kappa\,\eta_{\,t}\ +\ \div\bigl(\u_{\,0}\,\ue^{\,\kappa\,\eta}\bigr)\ =\ 0\,, \\
  \delta\u_{\,0}:&\quad \grad\Bigl((\div\u_{\,0})\,\ue^{\,2\,\kappa\,\eta}\Bigr)\ -\ \kappa^2\,\u_{\,0}\,\ue^{\,2\,\kappa\,\eta}\ =\ 2\,\kappa^3\,\phi_{\,0}\,\grad\eta\,\ue^{\,2\,\kappa\,\eta}\ -\ 2\,\kappa^2\,\grad\Bigl(\phi_{\,0}\,\ue^{\,2\,\kappa\,\eta}\Bigr)\,, \\
  \delta\eta:&\quad \bigl(\phi_{\,0}\,\ue^{\,\kappa\,\eta}\bigr)_{\,t}\ +\ \div\bigl[\,\phi_{\,0}\,\u_{\,0}\,\ue^{2\,\kappa\,\eta}\,\bigr]\ +\ g\,\eta\\
  &\quad -\ \kappa\,\div\Bigl(\u_{\,0}\,\ue^{\,\kappa\,\eta}\Bigr)\,\phi_{\,0}\,\ue^{\,2\,\kappa\,\eta}\ -\ \frac{1}{2}\;\Bigl[\,\abs{\u_{\,0}}^{\,2}\ +\ \frac{1}{\kappa^2}\;(\div\u_{\,0})^2\,\Bigr]\ =\ 0\,.
\end{align*}
Even if these equations are more compact (than the system we obtained in Section~\ref{sec:weak}) and the mass conservation coincides exactly with \eqref{eq:mass}, still this system seems to be quite complicated. This time, it follows from the variation $\delta\u_{\,0}$ that in order to reconstruct the horizontal velocity $\u_{\,0}$ from the velocity potential $\phi_{\,0}\,$, one has to solve an elliptic (vectorial) equation.

\bigskip
\paragraph*{Intermediate conclusions.} We saw above that the variational method can easily lead to complicated and unamenable equations, even if the latter inherits naturally the variational structure. Consequently, the choice of good ansatz is absolutely crucial for the derivation of an approximate model. In this respect a better ansatz will be proposed below.


\section{Alternative deep water ansatz}
\label{sec:deep}

\begin{figure}
  \centering
  \subfigure[$u_{\,1}\,(x_1,\,y,\,t)$]{\includegraphics[width=0.49\textwidth]{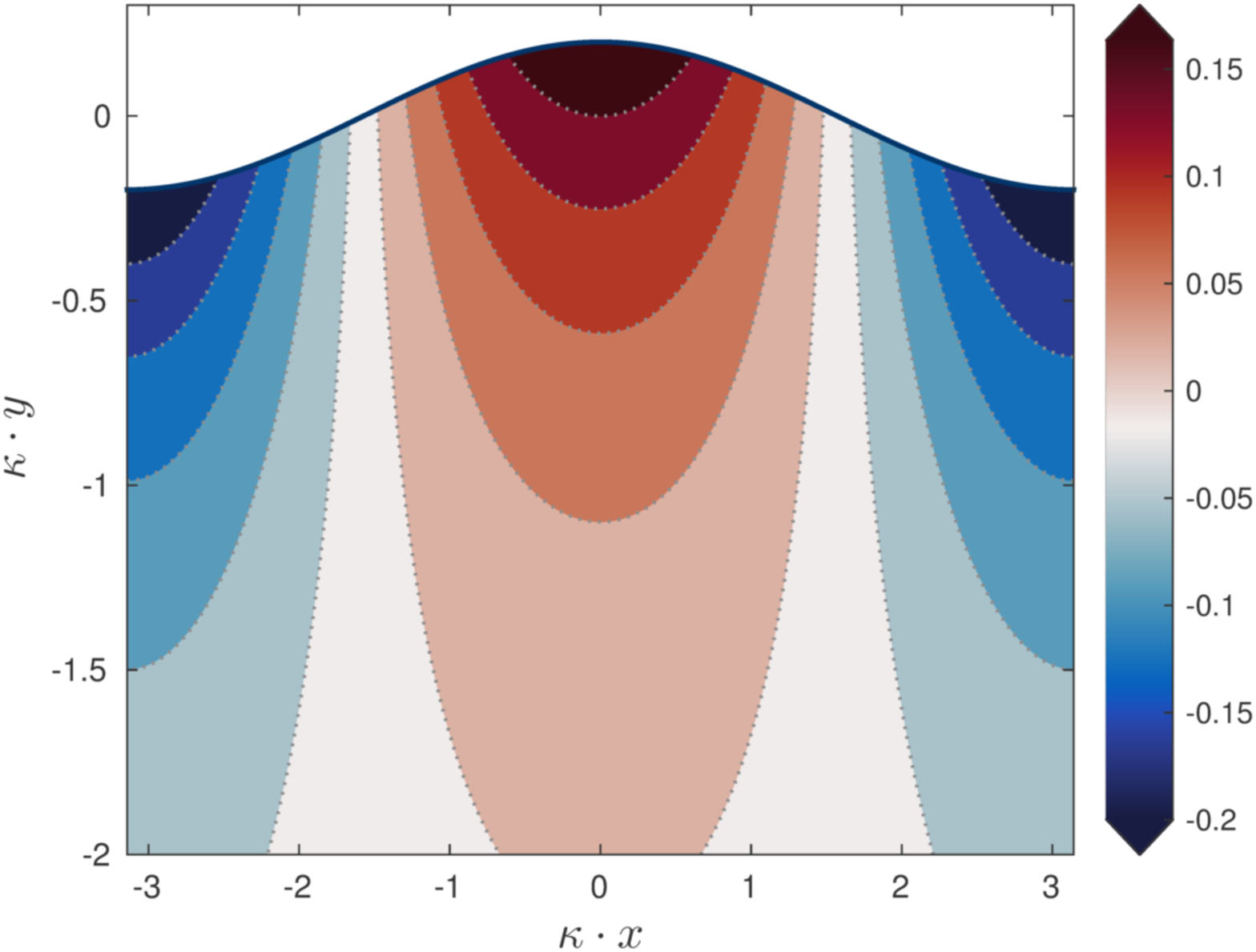}}
  \subfigure[$v\,(x_1,\,y,\,t)$]{\includegraphics[width=0.49\textwidth]{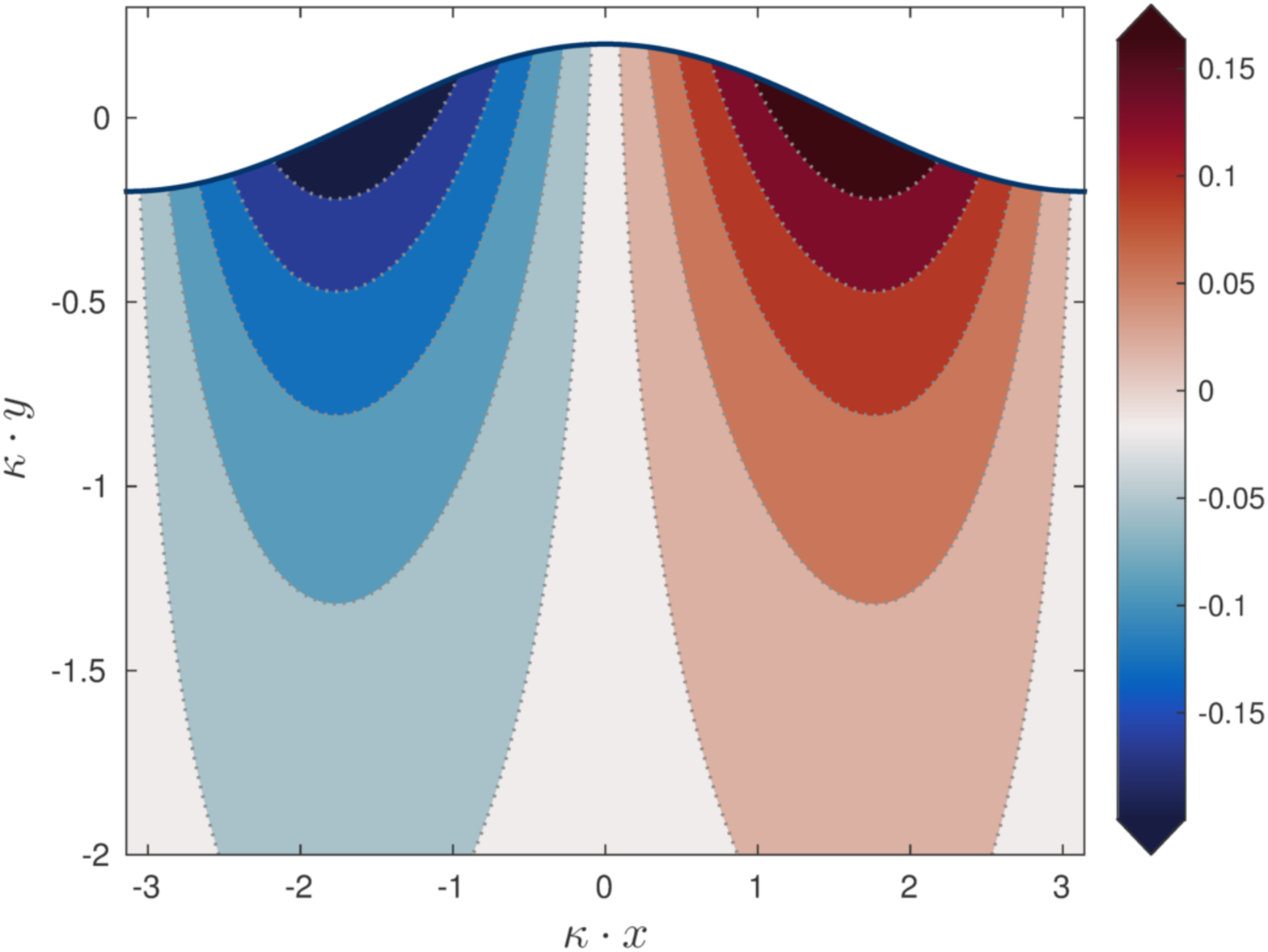}}
  \caption{\small\em Velocity field distribution predicted by ansatz \eqref{eq:ans2} under a periodic travelling wave: (a) horizontal and (b) vertical velocities. The periodic wave amplitude is $\alpha\ =\ 0.2\,$.}
  \label{fig:ans2}
\end{figure}

Let us slightly modify the ansatz \eqref{eq:ans1} in the following way: instead of defining all the quantities by their values at $y\ =\ 0\,$, we shall define them at the \emph{free surface} $y\ =\ \eta\,(\x,\,t)$ in accordance with \textsc{Hamilton}'s principle:
\begin{equation}\label{eq:ans2}
  \phi(\x,\,y,\,t)\, =\, \phis(\x,\,t)\;\ue^{\,\kappa\,(y\, -\, \eta)}\,, \ 
  \u(\x,\,y,\,t)\, =\, \ut(\x,\,t)\;\ue^{\,\kappa\,(y\, -\, \eta)}\,, \ 
  v(\x,\,y,\,t)\, =\, \vt(\x,\,t)\;\ue^{\,\kappa\,(y\, -\, \eta)}\,.
\end{equation}
The horizontal and vertical velocities predicted by this ansatz are represented in Figure~\ref{fig:ans2}. Notice qualitative similarities with Figure~\ref{fig:ans}. However, the magnitudes of velocities are slightly different.

Substituting the last ansatz into the relaxed variational principle \eqref{eq:relax} and performing exactly all integrations over the vertical coordinate $y$ yields the following \textsc{Lagrangian} density:
\begin{equation}\label{eq:Ldeep}
  2\,\kappa\,\L\ =\ 2\,\kappa\,\phis\,\eta_{\,t}\ -\ g\,\kappa\,\eta^2\ +\ \half\,\bigl(\abs{\ut}^{\,2}\ +\ \vt^{\,2}\bigr)\ -\ \ut\scal\bigl(\grad\phis\ -\ \kappa\,\phis\,\grad\eta\bigr)\ -\ \kappa\,\vt\,\phis\,.
\end{equation}
The direct comparison with \textsc{Lagrangian} \eqref{eq:incL} shows that just obtained \textsc{Lagrangian} density \eqref{eq:Ldeep} is much simpler. Indeed, in \eqref{eq:Ldeep} we have only cubic nonlinearities at most, while in \eqref{eq:incL} the nonlinearities are of infinite order. Below we shall derive some approximate models from this \textsc{Lagrangian} density in deep water.


\subsection{Saint-Venant equations in deep water}
\label{sec:sv}

The \textsc{Euler}--\textsc{Lagrange} equations for functional \eqref{eq:Ldeep} can be easily obtained:
\begin{align*}
  \delta\ut:&\quad \ut\ -\ \grad\phis\ +\ \kappa\,\phis\,\grad\eta\ =\ \vO\,, \\
  \delta\vt:&\quad \vt\ -\ \kappa\,\phis\ =\ 0\,, \\
  \delta\phis:&\quad 2\,\kappa\,\eta_{\,t}\ +\ \div\ut\ -\ \kappa\,\vt\ +\ \kappa\,\ut\scal\grad\eta\ =\ 0\,, \\
  \delta\eta:&\quad 2\,\kappa\,g\,\eta\ +\ 2\,\kappa\,\phis_{\,t}\ +\ \kappa\,\div(\phis\,\ut)\ =\ 0\,.
\end{align*}
The first two variations show that our approximation \eqref{eq:ans2} is \emph{exactly irrotational} in the sense that we have
\begin{equation*}
  \u\ \equiv\ \grad\phi\,, \qquad v\ \equiv\ \partial_{\,y}\,\phi\,.
\end{equation*}
The substitution of the first two relations in the last two yield the following system:
\begin{align*}
  \eta_{\,t}\ +\ \half\,\kappa^{\,-1}\,\grad^{\,2}\phis\ -\ \half\,\kappa\,\phis\ &=\ \half\,\phis\;\bigl[\,\grad^{\,2}\eta\ +\ \kappa\,\abs{\grad\eta}^{\,2}\,\bigr]\,, \\
  \phis_{\,t}\ +\ g\,\eta\ &=\ -\half\,\div\bigl[\,\phis\,\grad\phis\ -\ \kappa\,\phis^2\,\grad\eta\,\bigr]\,.
\end{align*}
The last system was called the \emph{generalized} \textsc{Klein}--\textsc{Gordon} \emph{equations} (gKG) since its linearization coincides with the classical \textsc{Klein}--\textsc{Gordon} \emph{equation}. The reasons why these equations can be considered as the analogue of nonlinear shallow water (or \textsc{Saint}-\textsc{Venant} \cite{SV1871}) equations in deep water are explained in \cite{Clamond2009}. The gKG equations were extensively studied in \cite{Dutykh2015a}. In particular, it was shown that these equations possess the canonical symplectic (\textsc{Hamiltonian}) and multi-symplectic structures additionally to the variational structure incorporated into the \textsc{Lagrangian} density \eqref{eq:Ldeep}. Moreover, it was shown numerically that gKG equations may develop waves with an angular point at the crest as it is known since G.~\textsc{Stokes} in the case of periodic travelling waves \cite{Stokes1880}. We note also that \textsc{Kraenkel} \etal \cite{Kraenkel2005} also obtained some indications for the existence of peaked travelling waves in their model in the limit of \emph{small aspect ratio} waves.

\begin{remark}
If we look for periodic linear travelling waves of the form
\begin{equation*}
  \eta(x\,,t)\ =\ \alpha\,\cos(k\,x\ -\ \omega\,t)\,,
\end{equation*}
then the dispersion relation will be
\begin{equation*}
  c_p(k)\ \equiv\ \frac{\omega(k)}{k}\ =\ \sqrt{\frac{1}{2}\;g\,\frac{k^2\ +\ \kappa^2}{\kappa\,k^2}}\,.
\end{equation*}
In particular, one can see that for $k\ =\ \kappa$ the phase and group velocities coincide with the exact values given by linearized \textsc{Euler} equations.
\end{remark}


\subsection{Serre equations in deep water}
\label{sec:dserre}

The next model of interest can be obtained if we impose the free surface impermeability as a constrain, \ie
\begin{equation*}
  \vt\ =\ \eta_{\,t}\ +\ \ut\scal\grad\eta\,.
\end{equation*}
By substituting this expression for $\vt$ into \textsc{Lagrangian} density \eqref{eq:Ldeep}, we obtain the following functional:
\begin{equation*}
  2\,\kappa\,\L\ =\ (\kappa\,\eta_{\,t}\ +\ \div\ut)\,\phis\ -\ \kappa\,g\,\eta^2\ +\ \half\,\abs{\ut}^{\,2}\ +\ \half\,(\eta_{\,t}\ +\ \ut\scal\grad\eta)^2\,.
\end{equation*}
The \textsc{Euler}--\textsc{Lagrange} equations give us the following relations among dependent variables:
\begin{align}\label{eq:serre1}
  \delta\ut:&\quad \ut\ +\ (\eta_{\,t}\ +\ \ut\scal\grad\eta)\ -\ \grad\phis\ =\ \vO\,, \\
  \delta\phis:&\quad \kappa\,\eta_{\,t}\ +\ \div\ut\ =\ 0\,,\label{eq:serre2} \\
  \delta\eta:&\quad \eta_{\,t\,t}\ +\ \kappa\,\phis_{\,t}\ +\ 2\,\kappa\,g\,\eta\ + \nonumber\\
  &\qquad (\ut\scal\grad\eta)_{\,t}\ +\ \div(\eta_{\,t}\,\ut)\ +\ \div\bigl[\,(\ut\scal\grad\eta)\,\ut\,\bigr]\ =\ 0\,.\label{eq:serre3}
\end{align}
The variation $\delta\phis$ shows that our approximation turns out to be exactly incompressible (since $\dfrac{\delta\,\L}{\delta\phis}\ =\ 0$ is equivalent to $\div\u\ +\ v_{\,y}\ =\ 0$). Notice that we did not require this condition at the level of ansatz \eqref{eq:ans2}. It is the \textsc{Hamilton} variational principle which gives this property automatically in this particular case. On the other hand, the approximation is not exactly irrotational. To linear approximation equations \eqref{eq:serre1} -- \eqref{eq:serre3} yield an improved linear \textsc{Boussinesq} equation:
\begin{equation*}
  (\grad^2\ -\ \kappa^2)\,\eta_{\,t\,t}\ +\ 2\,g\,\kappa\,\grad^2\,\eta\ =\ 0\,.
\end{equation*}
This \textsc{Boussinesq} equation admits travelling plane wave solutions with the dispersion relation
\begin{equation*}
  c_p(k)\ =\ \sqrt{\frac{2\,g\,\kappa}{k^2\ +\ \kappa^2}}\,.
\end{equation*}
Again, it is not difficult to see that for $k\ =\ \kappa$ we obtain the exact values of the phase and group velocities given by linear \textsc{Euler} equations. Notice, that to \emph{linear approximation} the model \eqref{eq:serre1} -- \eqref{eq:serre3} derived in this Section coincides with the model proposed by \textsc{Kraenkel} \etal \cite{Kraenkel2005} (see also Section~\ref{sec:models} above) if the modelling parameter $\m$ is chosen in the optimal way, \ie $\m\ =\ \kappa\,$. In \cite{Clamond2009} equations \eqref{eq:serre1} -- \eqref{eq:serre3} were named the deep water \textsc{Serre} equations (and this name was properly motivated). The existence of singular (\ie peaked) travelling waves was also shown in \cite{Clamond2009}.


\subsubsection{Evolution equations}

The governing equations \eqref{eq:serre1} -- \eqref{eq:serre3} may appear complicated to the reader. The main problem with the formulation given above is that equations are not written in an evolutionary form of a system of PDEs with time derivatives separated from other terms. We can recast equations \eqref{eq:serre1} -- \eqref{eq:serre3} in a more amenable form. In order to obtain a compact form of equations \eqref{eq:serre1} -- \eqref{eq:serre3}, first we are going to \emph{expand} them:
\begin{align*}
  \vt\ &\equiv\ \eta_{\,t}\ +\ \ut\scal\grad\eta\ =\ \frac{\eta_{\,t}\ +\ \grad\phis\scal\grad\eta}{1\ +\ \abs{\grad\eta}^{\,2}}\ \equiv\ \ut\scal\grad\eta\ -\ \kappa^{\,-1}\,\div\ut\,, \\
  \ut\ &\equiv\ \grad\phis\ -\ \vt\,\grad\eta\ =\ \frac{\grad\phis\ -\ \eta_{\,t}\,\grad\eta\ +\ \abs{\grad\eta}^{\,2}\,\grad\phis\ -\ \bigl[\,\grad\phis\scal\grad\eta\,\bigr]\,\grad\eta}{1\ +\ \abs{\grad\eta}^{\,2}}\,, \\
  0\ &=\ \kappa\,\eta_{\,t}\ +\ \div\ut\,, \\
  0\ &=\ 2\,\kappa\,g\,\eta\ +\ \kappa\,\phis_{\,t}\ +\ \vt_{\,t}\ +\ \div(\vt\,\ut)\,.
\end{align*}
The last equation gives us a hint that the right evolution variable is
\begin{equation*}
  \q\ \eqdef\ \grad\bigl(\phis\ +\ \kappa^{\,-1}\,\vt\bigr)\ =\ \ut\ +\ \vt\,\grad\eta\ +\ \kappa^{\,-1}\grad(\ut\scal\grad\eta)\ -\ \kappa^{\,-2}\,\grad(\div\ut)\,.
\end{equation*}
We can notice that $\grad(\div\ut)$ can be seen as an application of the operator matrix $\grad\otimes\grad$ to the vector $\ut$, \ie-
\begin{equation*}
  \q\ =\ \underbrace{\bigl[\,\Id\ -\ \kappa^{\,-2}\,\grad\otimes\grad\,\bigr]}_{\equiv\ \Dd^{\,-1}}\scal\,\ut\ +\ \vt\,\grad\eta\ +\ \kappa^{\,-1}\grad(\ut\scal\grad\eta)\,,
\end{equation*}
where $\Id$ denotes the identity operator. Finally, the system of evolution equations can be written as
\begin{align}\label{eq:dserre1}
  \kappa\,\eta_{\,t}\ +\ \div\Dd\scal\q\ &=\ \div\Dd\scal\bigl[\,\vt\,\grad\eta\ +\ \kappa^{\,-1}\,\grad(\ut\scal\grad\eta)\,\bigr]\,, \\
  \q_{\,t}\ +\ 2\,g\,\grad\eta\ &=\ -\kappa^{\,-1}\,(\grad\otimes\grad)\scal(\vt\,\ut)\,.\label{eq:dserre2}
\end{align}
These equations have to be supplemented by two algebro-differential relations:
\begin{align*}
  \vt\ &=\ \ut\scal\grad\eta\ -\ \kappa^{\,-1}\,\div\ut\,, \\
  \ut\ &=\ \Dd\scal\bigl[\,\q\ -\ \vt\,\grad\eta\ -\ \kappa^{\,-1}\grad(\ut\scal\grad\eta)\,\bigr]\,.
\end{align*}
The pseudo-differential operator $\Dd\ =\ \bigl[\,\Id\ -\ \kappa^{\,-2}\,\grad\otimes\grad\,\bigr]^{\,-1}$ can be easily computed in the \textsc{Fourier} space:
\begin{equation*}
  \hat{\Dd}\ =\ \bigl[\,\Id\ -\ \kappa^{\,-2}\,\k\otimes\k\,\bigr]^{\,-1}\,,
\end{equation*}
where $\k\ =\ (k_{\,1},\,k_{\,2})$ is the vector of wavenumbers. Otherwise, one has to invert an elliptic operator numerically as it is custom in the numerical analysis of classical \textsc{Serre} equations \cite{Dutykh2011a}.

\bigskip
\paragraph*{Open problem.}

By analogy to the classical \textsc{Serre}--\textsc{Green}--\textsc{Naghdi} equations the canonical \textsc{Hamiltonian} structure for deep water \textsc{Serre} equations \eqref{eq:dserre1}, \eqref{eq:dserre2} does not probably exist.
The Authors of the present manuscript did not succeed to find even a non-canonical \textsc{Hamiltonian} formulation which should exist in principle. Consequently, it remains an open problem so far. However, we succeeded to find the multi-symplectic formulation for deep water \textsc{Serre}-type equations.


\subsection{Multi-symplectic formulation}

The history of multi-symplectic formulations can be traced back to V.~\textsc{Volterra} (1890) who generalized \textsc{Hamiltonian} equations for variational problems involving several variables \cite{Volterra1890, Volterra1890a}. Later these ideas were developed in 1930's \cite{DeDonder1930, Weyl1935, Lepage1936}. Finally, in 1970's this theory was geometrized by several mathematical physicists \cite{Goldschmidt1973, Kijowki1974, Krupka1975, Krupka1975a} similarly to the evolution of symplectic geometry from the ideas of J.-L.~\textsc{Lagrange} \cite{Lagrange1853, Souriau1997}. In our study we will be inspired by modern works on multi-symplectic PDEs \cite{Bridges1997, Marsden1998}. Recently this theory has found many applications to the development of structure-preserving integrators \cite{Bridges2001, Moore2003a, Chen2011, Dutykh2013a}.

Here we give the multi-symplectic structure for deep water \textsc{Serre} equations in the case of one spatial horizontal dimension $x_{\,1}\ \equiv\ x$ (and, thus, $\tilde{u}\ \equiv\ \tilde{u}_{\,1}$) for the sake of notation compactness. The generalization to the case of two horizontal dimensions is straightforward. The general form of multi-symplectic equations (with one spatial variable) is
\begin{equation}\label{eq:ms}
  \M\scal \z_{\,t}\ +\ \Km\scal\z_{\,x}\ =\ \grad_{\z}\,\Ss\,(\z)\,,
\end{equation}
where $\z\ \in\ \R^{\,d}$ is the vector of state variables and $\M,\ \Km\ \in\ \Mat_{d\times d\,}(\R)$ are some \emph{skew-symmetric} matrices. It is not difficult to check that for deep water \textsc{Serre} equations \eqref{eq:serre1} -- \eqref{eq:serre3} (in 1D) it is sufficient to take
\begin{align*}
  \z\ &=\ {}^{\top}\bigl(\phis,\,\eta,\,\tilde{u},\,\gamma,\,\beta,\,\vt\bigr)\,, \\
  \Ss\,(\z)\ &=\ -g\,\kappa\,\eta^2\ +\ \beta\,(\tilde{u}\,\vt\ -\ \gamma)\ +\ \half\,\bigl(\tilde{u}^{\,2}\ -\ \vt^{\,2}\bigr)\,,
\end{align*}
and the skew-symmetric matrices $\M\,$ and $\Km$ are defined as
\begin{equation*}
  \M\ \eqdef\ \begin{pmatrix}
    0 & -\kappa & 0 & 0 & 0 & 0 \\
    \kappa & 0 & 0 & 0 & 0 & 1 \\
    0 & 0 & 0 & 0 & 0 & 0 \\
    0 & 0 & 0 & 0 & 0 & 0 \\
    0 & 0 & 0 & 0 & 0 & 0 \\
    0 & -1 & 0 & 0 & 0 & 0
  \end{pmatrix}\,, \qquad
  \Km\ \eqdef\ \begin{pmatrix}
    0 & 0 & -1 & 0 & 0 & 0 \\
    0 & 0 & 0 & 1 & 0 & 0 \\
    1 & 0 & 0 & 0 & 0 & 0 \\
    0 & -1 & 0 & 0 & 0 & 0 \\
    0 & 0 & 0 & 0 & 0 & 0 \\
    0 & 0 & 0 & 0 & 0 & 0
  \end{pmatrix}\,.
\end{equation*}
Equation \eqref{eq:ms} can be rewritten in the component-wise form for the sake of clarity:
\begin{align*}
  -\kappa\,\eta_{\,t}\ -\ \tilde{u}_{\,x}\ &=\ 0\,, \\
  \kappa\,\phis_{\,t}\ +\ \vt_{\,t}\ +\ \gamma_{\,x}\ &=\ -2\,g\,\kappa\,\eta\,, \\
  \phis_{\,x}\ &=\ \tilde{u}\ +\ \beta\,\vt\,, \\
  -\eta_{\,x}\ &=\ -\beta\,, \\
  0\ &=\ \gamma\ -\ \tilde{u}\,\vt\,, \\
  -\eta_{\,t}\ &=\ -\vt\ +\ \beta\,\tilde{u}\,.
\end{align*}
Now it is straightforward to check by making substitutions that equation \eqref{eq:ms} is indeed equivalent to the deep \textsc{Serre} equations \eqref{eq:serre1} -- \eqref{eq:serre3}. The generalization to the 2D case is straightforward.

\begin{remark}
The multi-symplectic structure for classical (\ie shallow water) one layer \textsc{Serre}--\textsc{Green}--\textsc{Naghdi} equations was recently reported in \cite{Chhay2016}. Later this structure was generalized to the case of interfacial waves between two layers (with rigid lid approximation) in \cite{Clamond2016a}.
\end{remark}


\section{Discussion}
\label{sec:disc}

Above we presented several approximate models in deep water regime and now we outline the main conclusions and perspectives of the present study.


\subsection{Conclusions}

In the present article first we discussed the most popular variational structures for the water wave problem in the deep water regime. Then, we assumed that the flow has an exponential profile in the vertical coordinate, which allowed us to derive some known and some new approximate models in deep water. The governing equations possess $\kappa-$dependent coefficients, where $\kappa$ is the dominant wave number that we try to describe in the physical domain. This property is quite usual for models describing the modulation of wave trains. However, in models we consider this property appears without introducing any envelopes. Whenever it was possible, the underlying variational structure of these equations was discussed. Of course, the best strategy to preserve this structure is to derive approximations using variational methods \cite{Clamond2009}. For instance, one can expand the \textsc{Hamiltonian} functional in powers of some order parameter and truncate this expansion after a few terms \cite{Zakharov1992} or one can choose an \emph{ansatz} for the flow, impose additional constraints, substitute everything into the \textsc{Lagrangian} density and perform simplifications \cite{Clamond2016}. The governing equations will be given by taking the variations of this functional with respect to all dependent variables. No matter which method is chosen, it is absolutely crucial to work and simplify the \textsc{Hamiltonian} or \textsc{Lagrangian} functionals instead of working with individual PDEs. In this study we illustrated also the weaknesses of the variational method. Indeed, a `bad' ansatz may lead to cumbersome and unamenable equations which have a variational structure, but nevertheless we do not really want to work with such model equations. Only the simultaneous application of the variational method to a `good' ansatz can yield new and practically useful approximate models.


\subsection{Perspectives}

Above we discussed mainly the modelling issues and we emphasized on the importance of the preservation of variational structure while deriving approximate models. However, these equations can be solved analytically only in rare special situations. Consequently, in the rest of cases we apply numerical numerical methods to solve these equations and we did not cover this topic in the present publication. For variational integrators in general we can refer to \cite{Lew2003}. However, these integrators are much better understood in the ODE (\ie symplectic) case. Some of the models given in this publication possess also the multi-symplectic formulation as well. The corresponding multi-symplectic numerical methods are available as well \cite{Bridges2001}. The general problem which exists in structure preserving numerical methods is to understand how symmetry and structure preservation affect usual properties of numerical methods (such as consistency, accuracy and stability) \cite{Chhay2011a, Chhay2008}. Of course, the quest for new interesting ans\"atze in deep water together with physically sound constraints has to be pursued.


\subsection*{Acknowledgments}
\addcontentsline{toc}{subsection}{Acknowledgments}

The authors would like to thank Professors V.~\textsc{Volpert} and V.~\textsc{Vougalter} for their kind invitation to prepare and submit this manuscript.
\bigskip


\addcontentsline{toc}{section}{References}
\bibliographystyle{abbrv}
\bibliography{biblio}

\end{document}